\documentclass{aa}  

\usepackage{graphicx}
\usepackage{txfonts}
\usepackage{morefloats}
\usepackage{hyperref}
\begin{document}

   \title{Partition functions 1: Improved partition functions and thermodynamic quantities for normal, equilibrium, and ortho and para molecular hydrogen\thanks{Tables \ref{tab:result-thermo-Qeq} to \ref{tab:result-thermo-Qpara} are only available in electronic form at the CDS via anonymous ftp to \texttt{cdsarc.u-strasbg.fr} (130.79.128.5) or via \url{http://cdsweb.u-strasbg.fr/cgi-bin/qcat?J/A+A/}}}
   \titlerunning{Partition functions and thermodynamic quantities of molecular hydrogen}

   \author{A. Popovas
          \inst{1}\fnmsep\thanks{Corresponding author.} 
          \and 
          ~ \medskip U. G. J\o rgensen\inst{1}
          }

   \institute{1. Niels Bohr Institute \& Centre for Star and Planet Formation, University of Copenhagen, \\
   \O ster Voldgade 5-7, 1350 Copenhagen K, Denmark
              \email{popovas@nbi.ku.dk}}

   \date{Received August 19, 2015; accepted July 1, 2016}

 
  \abstract
   {Hydrogen is the most abundant molecule in the Universe. Its thermodynamic quantities dominate the physical conditions in molecular clouds, protoplanetary disks, etc. It is also of high interest in plasma physics. Therefore thermodynamic data for molecular hydrogen have to be as accurate as possible in a wide temperature range.}
   {We here rigorously show the shortcomings of various simplifications
that are used to calculate the total internal partition function. These shortcomings can lead to errors of up to 40 percent or more in the estimated partition function. These errors carry on to calculations of thermodynamic quantities. Therefore a more complicated approach has to be taken.}
   {Seven possible simplifications of various complexity are described, together with advantages and disadvantages of direct summation of experimental values. These were compared to what we consider the most accurate and most complete treatment (case 8). Dunham coefficients were determined from experimental and theoretical energy levels of a number of electronically excited states of $\rm H_2$. Both equilibrium and normal hydrogen was taken into consideration.}
   {Various shortcomings in existing calculations are demonstrated, and the reasons for them are explained. New partition functions for equilibrium, normal, and ortho and para hydrogen are calculated and thermodynamic quantities are reported for the temperature range 1 - 20000 K. Our results are compared to previous estimates in the literature. The calculations are not limited to the ground electronic state, but include all bound and quasi-bound levels of excited electronic states. Dunham coefficients of these states of $\rm H_2$ are also reported.}
   {For most of the relevant astrophysical cases it is strongly advised to avoid using simplifications, such as a harmonic oscillator and rigid rotor or ad hoc summation limits of the eigenstates to estimate accurate partition functions and to be particularly careful when using polynomial fits to the computed values. Reported internal partition functions and thermodynamic quantities in the present work are shown to be more accurate than previously available data.}

   \keywords{Thermodynamic quantities --
                Hydrogen --
                Partition function --
                Dunham coefficients
               }

   \maketitle
%

\section{Introduction}

   The total internal partition function, $Q_{tot}(T)$, is used to determine how atoms and molecules in thermodynamic equilibrium are distributed among the various energy states at particular temperatures. It is the statistical sum over all the Boltzmann factors for all the bound levels. If the particle is not isolated, there is an occupation probability between 0 and 1 for each level depending on interactions with its neighbours. Together with other thermodynamic quantities, partition functions are used in many astrophysical problems, including equation of state, radiative transfer, dissociation equilibrium, evaluating line intensities from spectra, and correction of line intensities to temperatures other than given in standard atlases. Owing to the importance of $Q_{tot}(T)$, a number of studies were conducted throughout the past several decades to obtain more accurate values and present them in a convenient way. It is essential that a standard coherent set of $Q_{tot}(T)$ is being used for any meaningful astrophysical conclusions from calculations of different atmospheric models and their comparisons.

        Unfortunately, today we face a completely different situation. We have noted that most studies give more or less different results, sometimes the differences are small, but sometimes they are quite dramatic, for instance when different studies used different conventions to treat nuclear spin states (and later do not strictly specify these) or different approximations, cut-offs, etc. Furthermore, different methods of calculating the $Q_{tot}(T)$ are implemented in different codes, and it is not always clear which methods in particular are used. Naturally, differences in $Q_{tot}(T)$ values and hence in their derivatives (internal energy, specific heat, entropy, free energy) result in differences in the physical structure of computed model atmosphere even when line lists and input physical quantities (e.g. $T_{eff}$, log$g$, and metallicities) are identical.
        
        In the subsequent sections we review how $Q_{tot}(T)$ is calculated, comment on which simplifications, approximations, and constraints are used in a number of studies, and show how they compare to each other. We also argue against using molecular constants to calculate the partition functions. The molecular constants are rooted in the semi-classical idea that the
vibrational-rotational eigenvalues can be expressed in terms of a modified classical oscillator-rotor analogue. This erroneous concept has severe challenges at the highest energy levels that are not avoided by instead making a simple summation of experimentally determined energy levels, simply due to the necessity of naming the experimental levels by use of assigned quantum numbers. Instead, we report Dunham\cite{Dunham1932} coefficients for the ground electronic state and a number of excited electronic states, as well as resulting partition functions and thermodynamic quantities. The Dumham coefficients are not rooted in the classical picture, but still use quantum number assignments of the energy levels, and this approach is also bound to the same challenges  of defining the upper energy levels in the summation, as is the summation using molecular constants (and/or pure experimental data). No published studies have solved this challenge yet, but we quantify the uncertainty it implies on the resulting values of the chemical equilibrium partition function. 

\section{Energy levels and degeneracies}
In the Born-Oppenheimer approximation \cite{Born1927} it is assumed
that the rotational energies are independent of the vibrational energies, and the latter are independent of the electronic energies. Then the partition function can be written as the product of separate contributors - the external (i.e. translational) partition function, and the rotational, vibrational, and electronic partition functions. 
\subsection{Translational partition function} The $Q_{tr}$ can be expressed analytically as
\begin{equation}
Q_{tr} = \frac{Nk_BT}{\lambda^3 P},
\label{eq:q_tr}
\end{equation}
where
\begin{equation}
\lambda  = \sqrt{\frac{2\pi \hbar^2}{mk_BT}},
\end{equation}
and $N$ is the number of particles, $k_B$ is Boltzmann's constant, $P$ is the pressure, and $m$  the mass of the particle. The rest of this section considers the internal part, $Q_{int}$.
\subsection{Vibrational energies}
The energy levels of the harmonic oscillator (HO)
\begin{equation}
E(\rm v)=\left(\rm v+\frac{1}{2}\right)h\omega
\end{equation}
depend on the integer vibrational quantum number $\rm v = 0,1,2,.. .$ These energy levels are equally spaced by
\begin{equation}
\Delta E = h\omega.
\label{eq:vib-en-sep}
\end{equation}
The frequency $\omega = \frac{1}{2\pi}\sqrt{k/\mu}$ depends on the constant $k$, which is the force constant of the oscillator, and $\mu$, the reduced mass of the molecule. The lowest vibrational level is $E=\frac{1}{2}h\omega$. The harmonic oscillator potential approximates a real diatomic molecule potential well enough in the vicinity of the potential minimum at $R=R_e$, where $R$ is the internuclear distance and $R_e$ is the equilibrium bond distance, but it deviates increasingly for larger $\vert R-R_e \vert$. A better approximation is a Morse potential,
\begin{equation}
E_{pot}(R) = E_D\left[1-e^{-a(R-R_e)}\right]^2,
\end{equation}
where $E_D$ is the dissociation energy of the rigid molecule. The experimentally determined dissociation energy $E_D^{exp}$, where the molecule is dissociated from its lowest vibration level, has to be distinguished from the binding energy $E_B$ of the potential well, which is measured from the minimum of the potential. The difference is $E_D^{exp} = E_B - \frac{1}{2}h\omega$. The energy eigenvalues are
\begin{equation}
E(\rm v) = h\omega\left(\rm v + \frac{1}{2}\right)-\frac{h^2 \omega^2}{4E_D}\left(\rm v+\frac{1}{2}\right)^2
\end{equation}
with energy separations
\begin{equation}
\Delta E(\rm v) = E(\rm v+1)-E(\rm v) = h\omega\left[1-\frac{h\omega}{2E_D}(\rm v+1)\right].
\label{eq:vibration_energy_separations}
\end{equation}
When using a Morse potential, the vibrational levels are clearly
no longer equidistant: the separations between the adjacent levels decrease with increasing vibrational quantum number $\rm v$, in agreement with experimental observations. In astrophysical applications these energies are commonly expressed in term-values $G(\rm v) = E(\rm v)/hc,$ and the vibrational energies in the harmonic case are then given as
\begin{equation}
G(\rm v) =\omega_e \left(\rm v+\frac{1}{2}\right),
\end{equation}
and in the anharmonic case are given (when second-order truncation is assumed) as
\begin{equation}
G(\rm v) = \omega_e\left(\rm v+\frac{1}{2}\right)-\omega_ex_e\left(\rm v+\frac{1}{2}\right)^2
\label{eq:anharmonic}
,\end{equation}
where $\omega_e=\frac{\omega}{c}$ is the harmonic wave number, $\omega_ex_e=\frac{h \omega^2}{4cE_d}=\omega_e^2\frac{hc}{4E_D}$ is the anharmonicity term, and $\omega = a\sqrt{2E_D/\mu}$ is the vibrational frequency. We remark that literally, only $\omega$ is a frequency (measured in s$^{-1}$), while $\omega_e$ is a wave number (measured in cm$^{-1}$), but in the literature $\omega_e$ is commonly erroneously called a frequency any way. In the harmonic approximation, the Boltzmann factor (i.e. the relative population of the energy levels) can be summed analytically from $\rm v=0$ to $\rm v=\infty~$ \cite{Herzberg1960},
\begin{equation}
Q_{\rm v} \approx \sum_{\rm v=0}^\infty e^{-hc \omega_e(\rm v+\frac{1}{2})/k_BT} = (1-e^{-hc \omega_e/k_BT})^{-1}.
\label{eq:Herzberg1960}
\end{equation}

\subsection{Rotational energies}
In the rigid rotor approximation (RRA), a diatomic molecule can rotate around any axis through the centre of mass with an angular velocity $\omega_\angle$. Its rotation energy is
\begin{equation}
E(J) = \frac{1}{2}I\omega_\angle^2=\frac{J^2}{2I},
\end{equation}
here $I = \mu R^2$ is the moment of inertia of the molecule with respect to the rotational axis and $\vert J\vert=I\omega_\angle$ is its rotational angular momentum. In the simplest quantum mechanical approximation, the square of the classical angular momentum is substituted by $J(J+1)$,
\[\vert J\vert^2\rightarrow J(J+1)\hbar^2\]
and the rotational quantum number $J$ can take only discrete values. The rotational energies of a molecule in its equilibrium position are therefore represented by a series of discrete values
\begin{equation}
E(J) = \frac{J(J+1)\hbar^2}{2I},
\end{equation}
and the energy separation between adjacent rotational levels
\begin{equation}
\Delta E(J) = E(J+1)-E(J)=\frac{(J+1)\hbar^2}{2I}
\label{eq:rot-en-sep}
\end{equation}
is increasing linearly with J. Rotational term-values are
\begin{equation}
F(J) = \frac{J(J+1)h^2}{8\pi^2 I}=B_eJ(J+1),
\label{eq:rigrot}
\end{equation}
where
\begin{equation}
B_e=\frac{h^2}{8\pi^2cI}
\end{equation}
is the main rotational molecular constant and $F(J)$ is measured in the same units as $G(\rm v)$, that is, cm$^{-1}$. A mathematical advantage (but physically erroneous) of Eq. \ref{eq:rigrot} is that the rotational partition function $Q_{rot}$ can be calculated analytically,
\begin{equation}
Q_{rot} \approx \int (2J+1)e^{-hcB_eJ(J+1)/kT}dJ=\frac{kT}{hcB_e}.
\label{eq:rigrot_anal}
\end{equation}
A slightly more general approximation would be a non-rigid rotor. In this case, when the molecule rotates, the centrifugal force, $F_c = \mu\omega_\angle^2R \approx \mu\omega_\angle^2R_e$, acts on the atoms and the internuclear distance widens to a value $R$ that  is longer than $R_e$, and then the rotation energies are expressed as
\begin{equation}
E(J)\approx \frac{J(J+1)\hbar^2}{2I}-\frac{J^2(J+1)^2\hbar^4}{2k\mu^2R_e^6},
\label{eq:rotation_energies}
\end{equation}
where $k=4\pi^2\omega_\angle^2c^2\mu$ is the restoring force constant in a harmonic oscillator approximation, introduced above.\\
For a given value of the rotational quantum number $J$ the centrifugal widening makes the moment of inertia larger and therefore the rotational energy lower, as expressed in Eq. \eqref{eq:rotation_energies}. This effect compensates for the increase in potential energy. Using term-values, a centrifugal distortion constant $D_e$ is introduced into Eq. \eqref{eq:rigrot}, yielding
\begin{equation}
F(J) = B_eJ(J+1)-D_eJ^2(J+1)^2.
\label{eq:nonrig_rot}
\end{equation} 

 It is often assumed that two terms are enough to have a good approximation and that more terms only give a negligible effect on the partition function. Unfortunately, this is rarely the case. When higher vibrational and rotational states are taken into consideration, as we show in the subsequent sections, higher order terms must be used. The rotational partition function $Q_{rot}$ can be computed in the same way as the vibrational partition function. However, there are $2J+1$ independent ways the rotational axis can orient itself in space with the same given energy. Furthermore, a symmetry factor, $\sigma$, has to be added as well if no full treatment of the nuclear spin degeneracy is included. In this case, oppositely oriented homonuclear molecules are indistinguishable, and half of their $2J+1$ states are absent. Hence $\sigma$ is 2 for homonuclear molecules and unity for heteronuclear molecules, and this leads to half as many states in homonuclear molecules as in corresponding heteronuclear moleculs, such that Eq. \eqref{eq:rigrot_anal} becomes
 \begin{equation}
 Q_{rot} = \frac{kT}{\sigma h c B_e}.
\label{eq:sigma_qrot}
\end{equation}
In the full spin-split degeneracy treatment for homonuclear molecules, the degeneracy factor is given by
\begin{equation}
g_n = (2S_1+1)(2S_2+1)
\end{equation}
for the possible orientations of the nuclear spins $S_1$ and $S_2$. For the two nuclei, each with spin S, there are $S(2S+1)$ antisymmetric spin states and $(S+1)(2S+1)$ symmetric ones. For diatomic molecules, composed of identical Fermi nuclei\footnote{S=1/2, 3/2, 5/2, ...}, the spin-split degeneracy for even and odd $J$ states are
\begin{equation}
g_{n,even}=[(2S+1)^2+(2S+1)]/2,
\label{eq:spin}
\end{equation}
\[g_{n,odd}=[(2S+1)^2-(2S+1)]/2\]
respectively. The normalisation factor for $g_n$ is $1/(2S+1)^2$. For identical Bose nuclei\footnote{S=0, 1, 2, ...}, the spin-split degeneracy for even and odd $J$ states are opposite to what they are for Fermi systems.
\subsection{Interaction between vibration and rotation}
Interaction between vibrational and rotational motion can be allowed by using a different value of $B_{\rm v}$ for each vibrational level:
\begin{equation}
B_{\rm v} = B_e - \alpha_e (\rm v + \frac{1}{2}),
\label{eq:B_nu}
\end{equation}
where $\alpha_e$ is the rotation-vibration interaction constant.
\subsection{Electronic energies}
        The excited electronic energy levels typically are at much higher energies than the pure vibrational and rotational energy levels. For molecules they therefore contribute only a fraction to the partition function. Bohn \& Wolf \cite[1984]{Bohn1984} points out that for molecules the electronic state typically
is non-degenerate because the electronic energy is typically
higher than the dissociation energy, so that $Q_{el} = 1$. Although this is the case for most molecules, others do have degenerate states, therefore $Q_{el}$ is sometimes fully calculated nonetheless. Naturally, molecules can and do occupy excited electronic states, as many experiments show. The electronic partition function is expressed as
\begin{equation}
Q_{el} = \sum\tilde{\omega_e} e^{-T_e/k_BT},
\end{equation}
where $\tilde{\omega_e}=(2-\delta_{0,\Lambda})(2S+1)$ is the electronic statistical weight, where $\delta_{i,j}$ is the Kronecker delta symbol, which is 1 if $\Lambda = 0$ and 0 otherwise.
\subsection{Vibrational and rotational energy cut-offs and the basic concept of quantum numbers}

        If the semi-classical approach to the very high eigenstates (where this approximation has already lost its validity) is maintained
by assigning $\rm v$ and J quantum numbers either as reached from the Morse potential (Eq. \eqref{eq:vibration_energy_separations}) or from introducing second-order terms (as is most common), then the absurdity in the approach becomes more and more obvious. In Herzberg’s pragmatic interpretation \cite[, p. 99-102]{Herzberg1950} the classical theory is applied until the dissociation energy is reached (which for $H_2$ corresponds to v = 14), and then v values are disregarded above this. Other authors have not always been happy with this interpretation, however, and insisted on including v values well above the dissociation energy to a somewhat arbitrary maximum level, which affects the value of the partition function substantially at high temperatures.  In the Morse potential (Eq. \eqref{eq:vibration_energy_separations}) the vibrational eigenstates can in principle be counted up to v quantum numbers corresponding to approximately twice the dissociation energy, and for even higher v quantum numbers, the vibrational energy becomes a decreasing function of the increasing v number:
\begin{equation}
\Delta E(\rm v) = h\omega\left[ 1- \frac{h\omega}{2E_D}(\rm v+1)\right] < 0,
\end{equation}
solving for $\rm v$ where $\Delta E(\rm v)$ first becomes lower
than or equal to 0, we obtain
\begin{equation}
\rm v_{max} = \frac{2E_D-h\omega}{h\omega}.
\label{eq:vmax_break}
\end{equation}
The next level, $\rm v_{max}+1$, would produce zero or even negative energy addition. To solve this failure, more terms are needed to better approximate the real potential. Expressing $\rm v_{max}$ in vibrational term molecular constants, we obtain
\begin{equation}
\rm v_{max} = \frac{2E_D-h\omega}{h\omega} = \frac{\omega}{c}\frac{8cE_D}{2h\omega^2}=\frac{\omega_e}{2\omega_ex_e}-1.
\label{eq:v_max_long}
\end{equation}
$\rm v_{max}$ is sometimes used as a cut-off level for summation of vibrational energies (e.g. \cite{Irwin1987}).\newline
The classical non-rigid rotor approximation, Eq. \eqref{eq:nonrig_rot}
also shows that the energy between adjacent levels decreases with increasing J, thus writing Eq. \eqref{eq:nonrig_rot} for $J$ and $J+1$ states and making the difference between eigenstates equal to zero,
\begin{equation}
B_e(J+1)(J+2)-D_e(J+1)^2(J+2)^2-B_eJ(J+1)+D_eJ^2(J+1)^2=0,
\end{equation} 
and solving for $J$, we obtain the maximum rotational number,
\begin{equation}
J_{max} = \frac{\sqrt{2}\sqrt{B_e}-2\sqrt{D_e}}{2\sqrt{D_e}}.
\label{eq:Jmax_break}
\end{equation}
The summation over rotational energies is commonly cut-off at some arbitrary rotational level or is approximated by an integral from 0 to $\infty$. When examining the literature, we did not note this $J_{max}$ expression. As with the non-harmonic oscillator, adding more terms to the non-rigid rotor approximation would give a better approximation. Now it is possible to measure more terms ($\omega_ye_y, \omega_ze_z$, ... for vibrational levels, $H_e$ for rotational levels). Although they are important for high-temperature molecular line lists and partition functions, they are often calculated only from energy levels measured in the vicinity of the bottom of the potential well. For higher $\rm v$ and $J$ values, the resulting energies can therefore
still deviate from experimentally determined energy term values. Furthermore, the whole concept of quantum numbers $\rm v$ and $J$ and associated degeneracies etc. of course relates to the classical approximations in quantum theory, which breaks down at energies far from the validity of a description of the molecule as a rotating system of some kind of isolated atoms bound to one another by some kind of virtual spring. This shows that there is an unclear limit beyond which it is no longer meaningful to continue adding higher order terms to expand the simple quantum number concept to fit the highest measured energy levels. At these highest levels there is nothing other than the mathematical wavefunction itself that represent reality in a meaningful way.

\begin{figure}
   \centering
                {\includegraphics[width=\hsize]{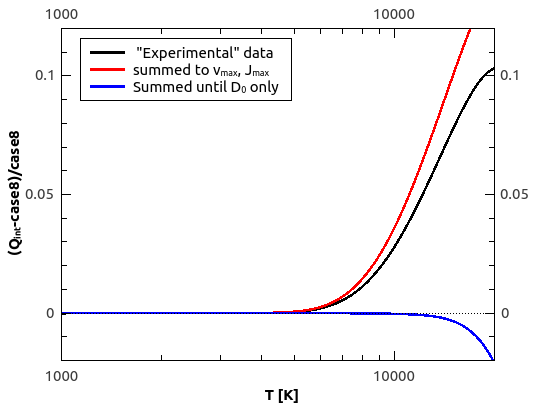}}
\caption{Difference between different methods used to stop the summation (upper cuts) of the partition function. Different cut-offs are compared to the partition function obtained from case 8 (see text for more details), which is the zero line. The curve marked “experimental “ data is derived by summing the Boltzmann factor of experimental data for the ground state used in \cite{Kurucz1985}\protect\footnotemark. The blue curve is is the summation for the ground electronic state summed to D$_0$ only. The red curve is the corresponding summation of energy levels of the three lowest electronic states based on Dunham coefficients and with a cut-off according to Eqs. \eqref{eq:mine_vMax} and \eqref{eq:mine_Jmax}.}
\label{fig:sum_check}
\end{figure}

Given these limitations in defining the uppermost levels in a meaningful way within the semi-classical quantum mechanics, the
question is whether in the end it is not more accurate to use only experimental energy levels in the construction of the partition function. However, exactly for the uppermost levels there is no such concept as experimentally determined energy levels summed to a partition function. The first problem is the practical one that for no molecule all the levels have been measured. For any missing level an assumption has to be applied, which could be to exclude the level in the summation, to make a linear fit of its energy between the surrounding measured energy levels, or to construct some type of polynomial fit from which to determine the missing levels, which would then be based on a choice of harmonic or anharmonic and rigid or non-rigid rotor or other polynomial fits. Another practical limitation is that experimentally, energy levels beyond the molecular dissociation energy would be seen. These levels can either give rise to a photo-dissociation of the molecule or relax back into a bound state, each with a certain probability. To construct a meaningful partition function, it would therefore at best have to be both temperature and pressure dependent, taking into account the timescales for molecular dissociation and recombination. An even more fundamental problem in defining a partition function based on experimental data is that each observed energy level has to be “named”. This naming is done by assigning a quantum number, which associates the measurements to a chosen semi-classical theory, however, and the quantum numbers associated with the state determines the degeneracy that the summation in the partition function will be multiplied
with. This dilemma becomes perhaps more clear for slightly larger molecules than $H_2$ where the degeneracy of the bending mode is multiplied directly into the partition function, but is a non-trivial function of the assigned vibrational quantum number of the state (\cite{Sorensen1993}), while at the same time quantum mechanical ab initio calculations show that the wavefunction of the uppermost states can only be expressed as a superposition of several wavefunctions that are expressed by each their set of quantum numbers (\cite{Jorgensen1985}), such that there is no “correct” assignment of the degeneracy factor that goes into the “experimental” partition function. There have been various ways in the literature to circumvent this problem. The most common is to ignore it. In some cases (e.g. \cite{Irwin1987} and \cite{Kurucz1985}) it was chosen to first complement the “experimental” values with interpolation between the measured values and then add extrapolated values above the dissociation energy (ignoring the pressure dependence) based on polynomial fits to the lower levels, with a cut-off of the added levels determined by where the order of the fit caused energy increments of further states to become negative (determined e.g. from Eqs. \eqref{eq:vmax_break} and \eqref{eq:Jmax_break}). For low temperatures these inconsistencies have no practical implication because the somewhat arbitrarily added eigenstates do not contribute to the final partition function within the noise level. For higher temperatures, however, the uncertainty due to the inconsistency of how to add energy levels above the dissociation energy becomes larger than the uncertainty in using a polynomial fit, rather than the actually computed values, as illustrated in Fig. \ref{fig:sum_check}. Realising that strictly speaking there is no such concept as an experimentally summed partition function, and there is no “correct” upper level, we decided for a hybrid model, where the uppermost energy level in our summations are the minimum of the last level before the dissociation energy and the level where the energy increments become negative according to Eqs. \eqref{eq:mine_vMax} and \eqref{eq:mine_Jmax}, see case 8 in the next section for more details. For high temperatures this approach will give results in between the results that would be obtained by a cut at the dissociation energy of the ground electronic state and an upper level determined by Eqs. \eqref{eq:mine_vMax} and \eqref{eq:mine_Jmax}, or a more experimentally inspired upper level (black curve), as is illustrated in Fig. \ref{fig:sum_check}. The difference between the three methods (upper cuts) is an illustration of the unavoidable uncertainty in the way the partition functions can be calculated today. In the future, computers may be powerful enough to make sufficiently accurate ab initio calculations of these uppermost energy levels (but today the upper part of the ab initio energy surface is most often fitted to the experiments rather than representing numerical results). 

\footnotetext{listed at \url{http://kurucz.harvard.edu/molecules/h2/eleroyh2.dat}}
        There are no restrictions on the number of energy eigenstates available for an isolated molecule before dissociation. However, the physical conditions of a thermodynamic system impose restrictions on the number of possible energy levels through the mean internal energy \cite{Cardona2005,Vardya1965}. A molecule in a thermodynamic system is subjected to the interaction with other molecules through collisions and with the background energy that permeates the system, making the number of energy states available dependent on the physical
conditions of the system. From a geometric point of view, in a thermodynamic system with total number density of particles $N$, the volume occupied by each molecule considering that the volume is cubic and of side $L$, is $L^3 = 1/N$. This imposes a maximum size that the molecule may have without invading the space of other molecules, and therefore the number of states becomes finite. From a physical point of view, the number of energy states of the molecules is delimited by the interactions with the rest of the system since the energy state above the last one becomes dissociated by the mean background thermal energy of the system under study. Hence the maximum number of states available for a particle is dependent on the properties of the surrounding system. \cite{Cardona2012} have shown that in a dense medium ($N \geq 10^{21}$ cm$^{-3}$) interaction between molecules and their surroundings starts to become important and that the physical approach dominates the geometrical one. They also derived equations for the maximum vibrational and rotational levels in such conditions, assuming the rigid rotor approximation.


\section{Simplifications used in other studies}
In this section we briefly review the simplifications in calculating $Q_{int}$ that have been used and how they compare to each other.
\subsection{Case 1.} This is the simplest case that is often given in textbooks: a simple product of harmonic oscillator approximation (HOA) and rigid rotor approximation (RRA), respectively summed and integrated to infinity, assuming Born-Oppenheimer approximation without interaction between vibration and rotation, and assuming that only the ground electronic state is present. $Q_{int}$ then becomes a simple product of Eqs. \eqref{eq:Herzberg1960} and \eqref{eq:v_max_long}:
\begin{equation}
Q_{int} = \frac{T}{c_2\sigma B_e}\frac{1}{1-exp(-c_2\omega_e/T)},
\label{eq:case1}
\end{equation}
where $c_2 = h c/k_B$ is the second radiation constant.
\subsection{Case 2.} This time, HOA is dropped. $\rm v_{max}$ is evaluated by Eq.  \eqref{eq:vmax_break}\footnote{Which has no physical meaning}, only ground electronic state and no upper energy limit are assumed:
\begin{equation}
\begin{split}
Q_{int} = exp\left[\frac{c_2}{T}\left(\frac{\omega_{e,1}}{2}-\frac{\omega_ex_{e,1}}{4}\right)\right]\\\frac{T}{\sigma c_2B_e}\sum_{\rm v=0}^{\rm v_{max}}exp\left[\frac{c_2}{T}\omega_e\left(\rm v+\frac{1}{2}\right)-\omega_ex_e\left(\rm v+\frac{1}{2}\right)^2\right],
\label{eq:case2}
\end{split}
\end{equation}
where the first term refers to the lowest vibrational level of the ground state instead of the bottom of the potential well.
\subsection{Case 3.} This is the same as case 2, but this time $B_e$ in Eq. \ref{eq:case2} is replaced by $B_{\rm v}$, obtained by Eq. \eqref{eq:B_nu} (i.e. the Born-Oppenheimer approximation is abandoned):
\begin{equation}
\begin{split}
Q_{int} = exp\left[\frac{c_2}{T}\left(\frac{\omega_{e,1}}{2}-\frac{\omega_ex_{e,1}}{4}\right)\right]\\ \sum_{\rm v=0}^{\rm v_{max}}\frac{T}{\sigma c_2B_{\rm v}}exp\left[\frac{c_2}{T}\omega_e\left(\rm v+\frac{1}{2}\right)-\omega_ex_e\left(\rm v+\frac{1}{2}\right)^2\right].
\label{eq:case3}
\end{split}
\end{equation}
\subsection{Case 4.} This is the same as case 3, but this time an arbitrary energy truncation for the vibrational energy is added. The most commonly used \cite{Tatum1966,Sauval1984,Rossi1985,Gamache1990,Goorvitch1994,Gamache2000,Fischer2003} value is $40000$ cm$^{-1}$:
\begin{equation}
\begin{split}
Q_{int} = exp\left[\frac{c_2}{T}\left(\frac{\omega_{e,1}}{2}-\frac{\omega_ex_{e,1}}{4}\right)\right]\\ \sum_{\rm v=0}^{\rm v_{max}=40000 ~cm^{-1}}\frac{T}{\sigma c_2B_{\rm v}}exp\left[\frac{c_2}{T}\omega_e\left(\rm v+\frac{1}{2}\right)-\omega_ex_e\left(\rm v+\frac{1}{2}\right)^2\right].
\label{eq:case4}
\end{split}
\end{equation}
\subsection{Case 5.} In this case, excited electronic states
are added, \begin{equation}
\begin{split}
Q_{int} = exp\left[\frac{c_2}{T}\left(\frac{\omega_{e,1}}{2}-\frac{\omega_ex_{e,1}}{4}\right)\right] \\ \sum_e^{T_{e,max}}\sum_{\rm v=0}^{\nu_{max}}\frac{\tilde{\omega_e}T}{\sigma c_2B_{\rm v}}exp\left[\frac{c_2}{T}\omega_e\left(\rm v+\frac{1}{2}\right)-\omega_ex_e\left(\rm v+\frac{1}{2}\right)^2+T_e\right],
\label{eq:case5}
\end{split}
\end{equation}
where $T_{e,max}=40000~ \rm cm^{-1}$.
\subsection{Case 6.} This case provides the full treatment: we finally drop the RRA. As in cases 3 - 5, this case also fully abandons the Born-Oppenheimer approximation because electronic
and vibrational and rotational states are interacting and depend on one another. Different vibrational and rotational molecular constants are used for the different electronic levels. Furthermore, the arbitrary truncation is dropped and the real molecular dissociation energy, $D_0$, is used for the ground state. In principle, molecules can occupy eigenstates above the dissociation energy, but they are not stable and dissociate almost instantly or relax to lower states below the dissociation level. Since the timescale for dissociation is usually shorter than for relaxation, molecules tend to dissociate rather than relax. Furthermore, a pre-dissociation state is usually present for molecules, where no eigenstates can be distinguished and a continuum of energies is present. This regime is beyond the scope of this work. For excited electronic states the ionisation energy is used instead of the dissociation energy. The equation for the full treatment takes the form
\begin{equation}
\begin{split}
Q_{int} = exp\left[\frac{c_2}{T}\left(\frac{\omega_{e,1}}{2}-\frac{\omega_ex_{e,1}}{4}\right)\right] \\ \sum_{e=0}^{T_{e,max}} \sum_{\rm v=0}^{\rm v_{max}}\sum_{J=0}^{J_{max}} \tilde{\omega_e}(2J+1)exp\left[\frac{-c_2}{T}(T_e+G_{\rm v}+F_J)\right],
\label{eq:case6}
\end{split}
\end{equation}
where $G_{\rm v}$ is calculated as in Eq. \eqref{eq:anharmonic}, $F_J = B_{\rm v} J(J+1) - D_\nu J^2(J+1)^2$, where $B_{\rm v}$ is calculated as in Eq. \eqref{eq:B_nu}, and $D_{\rm v} = D_e - \beta_e(\rm v+\frac{1}{2})$. $T_{e,max}$ is the term energy of the highest excited electronic state and is taken into account. For each electronic state the $\rm v$ and $J$ quantum numbers are summed to a maximum value that corresponds to the dissociation energy of that particular electronic state.

\subsection{Case 7.} For homonuclear molecules, nuclear spin-split degeneracy must also be taken into account. In this case, the form of the equation is
\begin{equation}
\begin{split}
Q_{int} = exp\left[\frac{c_2}{T}\left(\frac{\omega_{e,1}}{2}-\frac{\omega_ex_{e,1}}{4}\right)\right] \\ \sum_{e=0}^{e_{max}} \sum_{\rm v=0}^{\rm v_{max}}\sum_{J=0}^{J_{max}} \tilde{\omega_e}g_n(2J+1)exp\left[\frac{-c_2}{T}(T_e+G_{\rm v}+F_J)\right],
\label{eq:case7}
\end{split}
\end{equation}
where $g_n$ is calculated as in Eq. \eqref{eq:spin}.
\subsection{Case 8.} This is the most accurate model we present
here. Although cases 6 and 7 give reasonable results, they are still not sufficient at high temperatures. Molecular constants are conceptually rooted in classical mechanics (such as force constants of a spring and centrifugal distortion), and most often they are evaluated only from lower parts of the potential well and represent higher states poorly. For this reason it is more accurate to use a Dunham polynomial \cite{Dunham1932} to evaluate energy levels,
\begin{equation}
T_{Dun} = \sum_{i,k=0}^{i_{max},k_{max}} Y_{ik}(\rm v+\frac{1}{2})^i(J(J+1))^k,
\label{eq:dunham_like}
\end{equation}
where $Y_{ik}$ are Dunham polynomial fitting constants. These constants are not exactly related to classical mechanical concepts, such as bond lengths and force constants, and as such represent one step further into modern quantum mechanics. The lower order constants have values very close to the corresponding classical mechanics analogue constants used in cases 1-7, with the largest deviations occurring for H$_2$ and hydrides \cite[, p. 109]{Herzberg1950}. For example, the differences between the molecular constants of H$_2$ $X ~1s\sigma$ state (taken from \cite{Pagano2009}) and the corresponding Dunham coefficients, obtained in this work, are 0.12\% for $\omega_e$  and $Y_{10}$, 3\% for $\omega_ex_e$ and $Y_{20}$, 72\% for $\omega_ey_e$ and $Y_{30}$, 0.015\% for $B_e$ and $Y_{01}$, 1.8\% for $\alpha_e$ and $Y_{11}$, 5.9\% for $\gamma_e$ and $Y{21}$, 0.25\% for $D_e$ and $Y_{02}$, 40\% for $\beta_e$ and $Y_{12}$, etc. Dunham polynomial constants used together as a whole represent all experimental data more accurate than the classical picture in cases 1-7\footnote{In most cases the estimated energies from Dunham polynomials differ by less than 1 cm$^{-1}$ from experimentally determined energies. For high energies they can, however, sometimes differ by up to 20 cm$^{-1}$.}. Results from Eq. \eqref{eq:dunham_like} are then summed to obtain $Q_{int}$:
\begin{equation}
\begin{split}
Q_{int}=\sum_{e,\rm v,J}^{e_{max},\rm v_{max},J_{max}}\omega_e(2J+1)\frac{1}{8}[(2S+1)^2 \\ -(-1)^J(2S+1)]exp\left[ \frac{-c_2}{T}T_{Dun}\right].
\label{eq:case8}
\end{split}
\end{equation}
In this case, $\rm v_{max}$ and $J_{max}$ are evaluated numerically, defined as the first $\rm v$ or $J$, where $\Delta E \leq 0$ (as in case 2, but now based on less biased coefficients), with $\Delta E$ given as
\begin{equation}
\Delta E(\rm v) = \sum_{i=1}^{i_{max}}Y_{i,0}\left[\left(\rm v+\frac{3}{2}\right)^i-\left(\rm v+\frac{1}{2}\right)^i\right] \leq 0,
\label{eq:mine_vMax}
\end{equation}
for ${\rm v}_{max}$, and
\begin{equation}
\Delta E(J) = \sum_{i=0,k=1}^{i_{max},k_{max}}Y_{i,k}\left({\rm v}+\frac{1}{2}\right)^i\left[(J+1)^k(J+2)^k-J^k(J+1)^k\right] \leq 0
\label{eq:mine_Jmax}
\end{equation}
for $J_{max}$.

\section{Thermodynamic properties}
From $Q_{tot}$ and its first and second derivatives we calculate thermodynamic properties. For $Q_{tr}$ (Eq. \eqref{eq:q_tr}) we assume 1 mol of ideal-gas  molecules, $N=N_A$, Avogadro's number. The ideal-gas standard-state pressure (SSP) is taken to be $p^\circ = 1$ bar and the molecular mass is given in amu.  The internal contribution $E_{int}$ to the thermodynamic energy, the internal energy $U-H(0)$, the enthalpy $H-H(0)$, the entropy $S$, the constant-pressure specific heat $C_{p}$, the constant-volume specific heat $C_{v}$, the Gibbs free energy $G-H(0),$ and the adiabatic index $\gamma$ are obtained respectively from
\begin{equation}
E_{int} = RT^2\frac{\partial lnQ_{int}}{\partial T}
\end{equation}
\begin{equation}
U-H(0)= E_{int}+\frac{3}{2}RT
\end{equation}
\begin{equation}
H-H(0)= E_{int}+\frac{5}{2}RT
\end{equation}
\begin{equation}
S = R~ lnQ_{tot} + \frac{RT}{Q_{tot}}\left( \frac{\partial Q_{tot}}{\partial T} \right)
\end{equation}
\begin{equation}
C_p=R\left[ 2T \left(\frac{\partial lnQ_{int}}{\partial T}\right)+T^2\left(\frac{\partial^2 ln Q_{int}}{\partial T^2}\right)\right]+\frac{5}{2}R
\end{equation}
\begin{equation}
C_v=\frac{\partial E_{int}}{\partial T} +\frac{3}{2}R
\end{equation}
\begin{equation}
G-H(0) = -RT~lnQ_{tot}+RT
\end{equation}
\begin{equation}
\gamma=\frac{H-H(0)}{U-H(0)}
.\end{equation}


\section{Orto/para ratio of molecular hydrogen}
In addition to equilibrium hydrogen, we also calculate the \emph{\textup{normal}} hydrogen following the paradigm of \cite{Colonna2012}. In this case, ortho-hydrogen and para-hydrogen are not in equilibrium, but act as separate species. This aspect is particularly important for simulations of protoplanetary disks. The equilibrium timescale is short enough (300 yr) that the ortho/para ratio (OPR) can thermalise in the lifetime of a disk, but the equilibrium timescale is longer than the dynamical timescale inside about 40 AU \cite{Boley2007}. In case of normal hydrogen, ortho and para states have to be treated separately. Then Eq. \eqref{eq:case8} is split into two, one for para states:
\begin{equation}
Q_{para}=\sum_{e,\rm v,J_{even}}^{e_{max},\rm v_{max},J_{max}}\omega_e(2J+1)exp\left[ \frac{-c_2}{T}T_{Dun}\right],
\label{eq:normalH2_para}
\end{equation}
and one for orhto states:
\begin{equation}
Q_{ortho}=\sum_{e,\rm v,J_{odd}}^{e_{max},\rm v_{max},J_{max}}\omega_e(2J+1)exp\left[ \frac{-c_2}{T}(T_{Dun}-\varepsilon_{001})\right].
\label{eq:normalH2_ortho}
\end{equation}
We note that weights for nuclear spin-split are omitted. The partition function for ortho-hydrogen is (as opposed to $Q_{para}$) referred to the first excited rotational level ($J=1$) of the ground electronic state in its ground vibrational level. $\varepsilon_{001}$ can be considered as the formation energy of the ortho-hydrogen
\cite{Colonna2012}.  $Q_{int}$ is then \cite{Colonna2012}
\begin{equation}
Q_{int}^{norm}=(Q_{para})^{g_p}(Q_{ortho})^{g_o},
\end{equation}
where $g_p$ and $g_o$ are the statistical weights of the ortho and para configurations, $0.25$ and $0.75,$ respectively. Moreover, the OPR is not necessarily either 3 or in equilibrium, but can be a time-dependant variable. The last consideration is beyond the scope of this work, therefore we do not discuss different ratios and assume that there are only two possibilities: either $\rm H_2$ is in equilibrium or not.


\section{Results and comparison}
        We finally are ready to compare these different approaches. We here focus on  $\rm H_2$, but we plan to apply the theory given above to other molecules in a forthcoming paper. Since $Q_{int}$ varies with several orders of magnitude as a function of temperature, we normalised the differences to case 8 : $\frac{Q_{tot,X}-Q_{tot,8}}{Q_{tot,8}}$, where $X$ is the case number.

        Molecular constants for cases 1-7 were taken from the NIST\footnote{K.P. Huber and G. Herzberg, "Constants of Diatomic Molecules" (data prepared by J.W. Gallagher and R.D. Johnson, III) in NIST Chemistry WebBook, NIST Standard Reference Database Number 69, Eds. P.J. Linstrom and W.G. Mallard, National Institute of Standards and Technology, Gaithersburg MD, 20899, http://webbook.nist.gov, (retrieved July 19, 2014).} database, summarised in \cite{Pagano2009}. $D_0=36118.0696$ cm$^{-1}$ \cite{Pisz2009}. 
Experimental and theoretical data for molecular hydrogen for case 8 was taken from a number of sources that are summarised in Table \ref{tab:H2data}. Theoretical calculations were used only when observational data were incomplete. For every electronic state, coefficients were fitted to all $\rm v-J$ configurations \emph{\textup{simultaneously}} using a weighted Levenberg-Marquardt \cite{Transtrum2012} algorithm and giving lower weights to theoretical data.

\begin{figure*}
   \centering
                {\includegraphics[width=17cm]{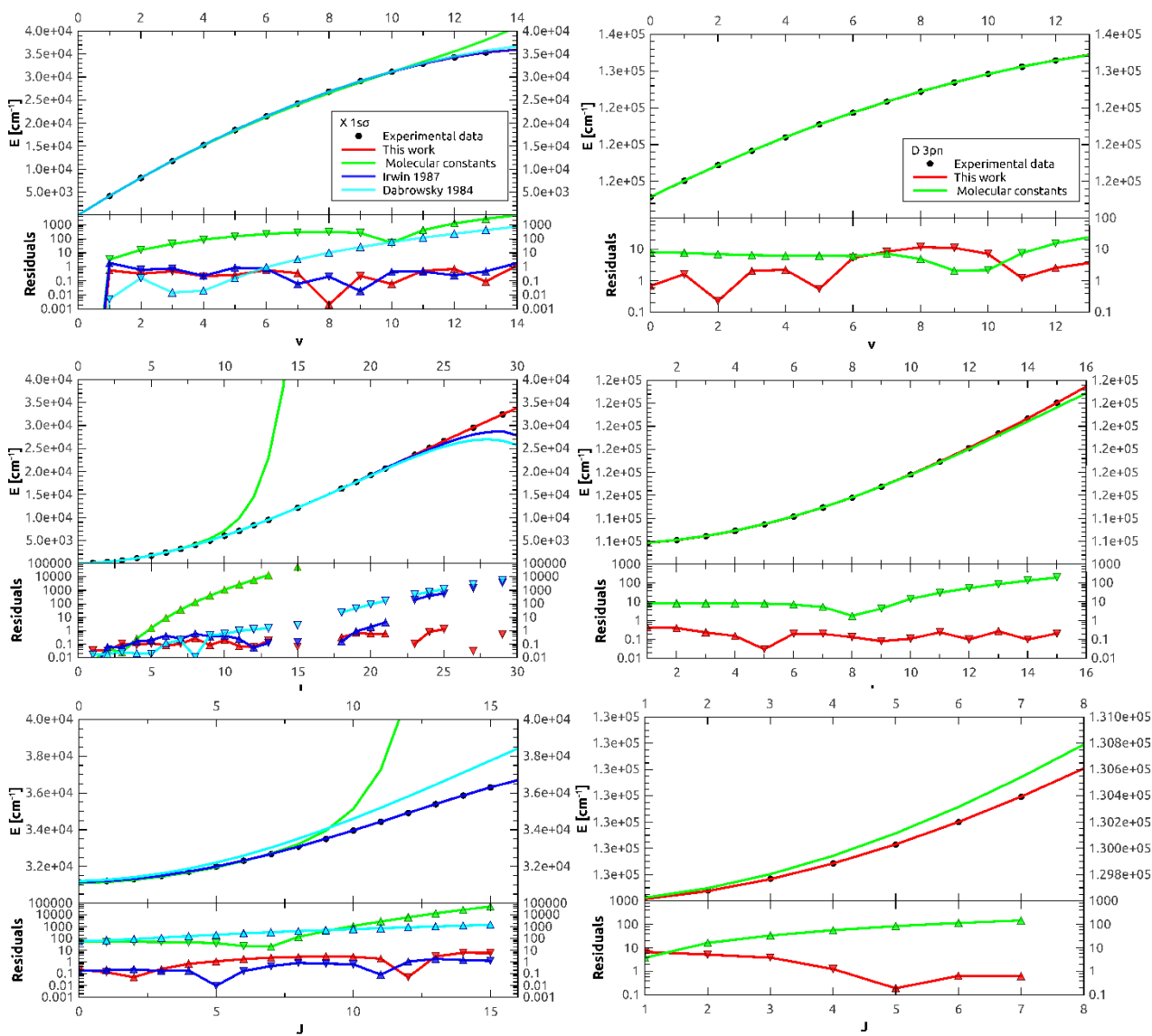}}
\caption{Energy levels for the $X~ 1s\sigma$ and $D~ 3p\pi$ states of $\rm H_2$. Top left: $X~ 1s\sigma$ vibrational energies. Top right: $D~ 3p\pi$ vibrational energies. Middle left: $X~ 1s\sigma ~(\rm v=0)$ energy levels. Middle right: $D~ 3p\pi ~(\rm v=0)$ energy levels. Bottom left: $X~ 1s\sigma ~(\rm v=10)$ energy levels. Bottom right: $D~ 3p\pi ~(\rm v=10)$ energy levels. See text for more details.}
\label{fig:energy_levels}
\end{figure*}

\subsection{Determination of Dunham coefficients and evaluation of their accuracy}
In this subsection we show how estimated energies from our Dunham coefficients  for particular $e,\rm v,J$ configurations compare to those estimated from molecular constants and to experimental data. For simplicity we will here discuss only two electronic ($X~ 1s\sigma$, $D~ 3p\pi$)\footnote{These states have a sufficiently large amount of molecular constants determined. Most other states have only few to no molecular constants determined, thus comparing them would be ungentlemanly.} and vibrational ($\rm v=$0 and 10) configurations of molecular hydrogen. Additionally, for the $X~ 1s\sigma$ state we show Dunham coefficients from \cite{Dabrowski1984} and \cite{Irwin1987}. For \cite{Dabrowski1984} signs had to be inverted for the $Y_{02}$, $Y_{04}$, $Y_{12}$, $Y_{14}$ , and $Y_{22}$ coefficients so that they would correctly follow the convention presented in Eq. \eqref{eq:dunham_like}. The comparison is shown in Fig. \ref{fig:energy_levels}. The top panels show vibrational energies of both states, middle panels
show energy levels of the $\rm v=0$ vibrational state, and bottom panels show energy levels of the $\rm v=10$ vibrational state. Experimental data are shown as black dots, red curves represent calculations of this work, green curves calculations from molecular constants, blue and cyan calculations using Dunham coefficients from \cite{Irwin1987} and \cite{Dabrowski1984}, respectively. The top panels in all panels show the resulting energies, the
bottom panels the residuals (calculated - experimental) in logarithmic scale. Triangles, facing down, indicate that the calculated energy is lower than observed, upward facing triangles that it is higher. These figures show that energies calculated from molecular constants can diverge substantially, especially in the $X~ 1s\sigma$ state, which contributes most to $Q_{int}$. For most of the states the results of \cite{Irwin1987} are of similar quality as ours (but we used fewer coefficients, which speeds up the $Q_{int}$ calculation), while for the lower vibrational levels of the $X~ 1s\sigma$ state our results are much more accurate (especially at high $J$).
        A complete set of Dunham coefficients is given in Tables \ref{tab:dunham-1ssigma} to \ref{tab:dunham-7ppi} (available online). The number of coefficients for all cases was chosen separately from examining the RMSE (root mean square error, compared to experimental/theoretical data) to determine those with the
lowest number of coefficients and lowest RMSE. In some cases as little as a 5x5 matrix is enough (e.g. $EF ~~ 2p\sigma 2p\sigma^2$), whereas in other cases up to a 11x11 matrix is needed. These differences depend on the complexity of the shape of the potential well. In all the tables the limiting energies are given up to which particular matrices should be used. All the limitations have to be employed (maximum energy as well as Eqs. \eqref{eq:mine_vMax} and \eqref{eq:mine_Jmax}) for every electronic configuration, since all the coefficients are only valid under these limitations.

\begin{table*}[h]
\footnotesize
\caption{Summary of data sources for $\rm H_2$ energy levels}
\centering
\begin{tabular}{c|p{6cm} p{6cm}} 
\hline
State & Experimental (levels) & Theoretical (levels) \\ 
\hline
\hline
$X~~ 1s\sigma$ & \cite{Dabrowski1984},($\rm v=0-14, J=0-29$) & \\
\hline
$B~~ 2p\sigma$ & \cite{Dabrowski1984},($\rm v=0-17, J=0-30$) \newline \cite{Abgrall1993}, ($\rm v=0-33, J=0-28$) \newline \cite{Bailly2010}, ($\rm v=0-13, J=0-13$)  & \cite{Wolniewicz2006}, ($\rm v=0-38, J=0-10$) \\
\hline
$C~~ 2p\pi$ & \cite{Dabrowski1984},($\rm v=0-13, J=1-13$) \newline \cite{Abgrall1993}, ($\rm v=0-13, J=1-24$) \newline \cite{Bailly2010}, ($\rm v=0-3, J=0-7$) & \cite{Wolniewicz2006}, ($\rm v=0-13, J=1-10$) \\
\hline
$EF^{[a]}~~ 2p\sigma 2p\sigma^2$ & \cite{Senn1987},($\rm v=0-20, J=0-5$) \newline \cite{Bailly2010}, ($\rm v=0-28, J=0-13$) & \cite{Ross1994}, ($\rm v=0-28, J=0-5$) \\
\hline
$B'~~ 3p\sigma$ & \cite{Abgrall1994}, ($\rm v=0-7, J=0-12$)\newline \cite{Bailly2010}, ($\rm v=0-3, J=0-7$) & \cite{Wolniewicz2006}, ($\rm v=0-9, J=0-10$) \\
\hline
$D~~ 3p\pi$ & \cite{Abgrall1994}, ($\rm v=0-14, J=1-17$) \newline \cite{Bailly2010}, ($\rm v=0-2, J=1-7$) & \cite{Wolniewicz2006}, ($\rm v=0-17, J=1-10$) \\
\hline
$GK^{[a]}~~ 3d\sigma$ & \cite{Bailly2010}, ($\rm v=0-6, J=0-9$) & \cite{Ross1994}, ($\rm v=0-5, J=0-5$)\newline \cite{Yu1994} ($\rm v=0-8,J=0-5$) \\
\hline
$H\overline{H}^{[a]}~~ 3s\sigma$ &\cite{Bailly2010}, ($\rm v=0-2, J=0-6$) &  \\
\hline
$I~~ 3d\pi$ &\cite{Bailly2010}, ($\rm v=0-3, J=1-10$) &  \\
\hline
$J~~ 3d\delta$ &\cite{Bailly2010}, ($\rm v=0-2, J=2-6$) &  \\
\hline
$B"Bbar^{[a]}~~ 4p\sigma$ & \cite{Glass2007},($\rm v=4-11, J=0-2$) & \cite{Wolniewicz2006}, ($\rm v=0-56^{[b]}, J=0-10$) \newline \cite{Glass2007},($\rm v=4-11, J=0-2$) \\
\hline
$D'~~ 4p\pi$ & \cite{Takezawa1970}, ($\rm v=0-5, J=1-5$) & \cite{Wolniewicz2006}, ($\rm v=0-9, J=1-10$)  \newline \cite{Glass2009},($\rm v=0-17, J=1-4$) \\
\hline
$5p\sigma$ & \cite{Takezawa1970}, ($\rm v=0-2, J=0-4$) & \cite{Wolniewicz2006}, ($\rm v=0-9, J=0-10$)  \newline \cite{Glass2013},($\rm v=0-8, J=0-4$) \\
\hline
$D" ~~ 5p\pi$ & \cite{Takezawa1970}, ($\rm v=0-2, J=1-3$) \newline \cite{Glass2013b},($\rm v=0-12, J=1-4$) & \cite{Glass2013b},($\rm v=0-12, J=1-4$) \\
\hline
$6p\sigma$ & \cite{Takezawa1970}, ($\rm v=0-2, J=0-4$) & \cite{Wolniewicz2006}, ($\rm v=0-9, J=0-10$)  \newline \cite{Glass2013},($\rm v=0-7, J=0-4$) \\
\hline
$6p\pi$ &\cite{Glass2013c},($\rm v=0-9, J=1-4$) &  \\
\hline
$7p\sigma$ & \cite{Takezawa1970}, ($\rm v=0-2, J=0-3$) & \cite{Glass2013},($\rm v=0-6, J=0-4$) \\
\hline
$7p\pi$ & \cite{Glass2013c},($\rm v=0-7, J=1-4$) &   \\
\hline
\end{tabular}
\tablefoot{$^{[a]}$Double potential well state. $^{[b]}$32 lowest $\rm v$ levels are used in this study.}
\label{tab:H2data}
\end{table*}

\onltab{
\begin{table*}[h]
\scriptsize
\caption{$Y_{i,j}$ for $X~ 1s\sigma$ state.  $E_{max}=E_D=36118.0696$.}
\begin{tabular}{c|ccccccc} 
\hline
$i \backslash j$ & 0&1&2&3&4&5&6\\ 
\hline 
0  & 0.0 & 60.8994 & -0.0464547 & 4.6066e-5 & -4.44761e-8 & 2.89037e-11 & -8.51258e-15\\
1  & 4408.97 & -3.22767 & 0.00251022 & -2.85502e-6 & 2.15163e-9 & -7.60784e-13\\      
2  & -127.648 & 0.165697 & -0.000458971 & 5.036e-7 & -2.1465e-10\\                  
3  & 2.90163 & -0.031327 & 6.61024e-5 & -4.15522e-8\\                             
4  & -0.302736 & 0.00278106 & -3.45438e-6\\                                     
5  & 0.0175198 & -0.000105554\\                                               
6  & -0.000606749\\
\hline 
\end{tabular}
\label{tab:dunham-1ssigma}
\end{table*}}

\onltab{
\begin{table*}[h]
\scriptsize
\caption{$Y_{i,j}$ for $B~~ 2p\sigma$ state. $E_{max}=118376.981$.}
\begin{tabular}{c|cccccccc} 
\hline
$i \backslash j$ & 0&1&2&3&4&5&6&7\\ 
\hline 
0  & 89540.7 & 20.3249 & -0.0244189 & 6.10056e-5 & -1.18524e-7 & 1.20211e-10 & -4.73402e-14 & 2.11275e-20 \\
1  & 1323.2 & -1.09933 & 0.00881566 & -4.47483e-5 & 1.02297e-7 & -1.06706e-10 & 4.16538e-14 & -1.58919e-19 \\             
2  & 1.11679 & -0.153064 & -0.000546492 & 6.67274e-6 & -1.75266e-8 & 1.8647e-11 & -7.16464e-15 & 4.0682e-20 \\                      
3  & -4.13726 & 0.063354 & -0.000205483 & 7.37966e-8 & 4.55652e-10 & -6.51076e-13 & 2.62926e-16 & -2.00198e-21\\                                 
4  & 0.394415 & -0.0062084 & 2.25521e-5 & -2.5732e-8 & 1.48268e-12 & 2.0649e-14 & -1.16798e-17\\                                               
5  & -0.0155583 & 0.000194542 & -5.03613e-7 & 1.07722e-10 & 4.69293e-13 & -4.60339e-16\\                                                     
6  & 8.69305e-5 & 3.59161e-6 & -2.00997e-8 & 3.79657e-11 & -1.75735e-14\\                                                                  
7  & 1.09635e-5 & -3.72309e-7 & 9.95017e-10 & -7.35739e-13\\                                                                             
8  & -2.49984e-7 & 8.41556e-9 & -1.11877e-11\\                                                                                         
9  & -6.45466e-11 & -6.35193e-11\\                                                                                                   
10  & 3.13732e-11\\ 
\hline
\hline
$i \backslash j$ &8&9&10\\
\hline
0& -2.50566e-19 & -2.92577e-19 & 1.61127e-19\\
1& 7.7568e-20 & 2.10342e-20\\ 
2& -2.63039e-21\\ 
\hline 
\end{tabular}
\label{tab:dunham-2psigma}
\end{table*}}

\onltab{
\begin{table*}[h]
\scriptsize
\caption{$Y_{i,j}$ for $C~~ 2p\pi$ state. $E_{max}=118376.981$.}
\begin{tabular}{c|cccccccc} 
\hline
$i \backslash j$ & 0&1&2&3&4&5&6&7\\ 
\hline 
0  & 97882.0 & 32.5557 & -0.0399421 & 0.000132418 & -3.34989e-7 & 4.2892e-10 & -2.12309e-13 & 3.53673e-19\\
1  & 2446.06 & -3.09621 & 0.0222126 & -0.000114815 & 2.84875e-7 & -3.3416e-10 & 1.48677e-13 & -1.64883e-19\\            
2  & -68.1303 & 0.685829 & -0.00833825 & 3.73323e-5 & -7.62593e-8 & 7.06128e-11 & -2.30836e-14\\                      
3  & -0.0313063 & -0.150771 & 0.00152702 & -5.22981e-6 & 7.37665e-9 & -3.65104e-12\\                                
4  & 0.0746035 & 0.019201 & -0.000146625 & 3.3335e-7 & -2.36923e-10\\                                             
5  & 0.00144141 & -0.00146403 & 7.19968e-6 & -8.04132e-9\\                                                      
6  & -0.00185288 & 6.26612e-5 & -1.45669e-7\\                                                                 
7  & 0.000162238 & -1.20807e-6\\                                                                            
8  & -4.75062e-6\\ 
\hline
\hline
$i \backslash j$ &8\\
\hline
0&  -1.42451e-20\\
\hline 
\end{tabular}
\label{tab:dunham-2ppi}
\end{table*}}

\onltab{
\begin{table*}[h]
\scriptsize
\caption{$Y_{i,j}$ for $EF~~ 2p\sigma 2p\sigma^2$ state, inner well. $E_{max}=105000.0$.}
\begin{tabular}{c|ccccc} 
\hline
$i \backslash j$ & 0&1&2&3&4\\ 
\hline 
0  & 98068.8 & 23.2211 & 2.25688 & -0.149909 & 0.00273895\\
1  & 2136.55 & -8.10192 & 0.679069 & -0.00863188\\       
2  & 140.173 & -1.30784 & -0.14758\\                   
3  & 9.85717 & 1.4317\\                              
4  & -22.7252\\
\hline 
\end{tabular}
\label{tab:dunham-EFinner}
\end{table*}}

\onltab{
\begin{table*}[h]
\scriptsize
\caption{$Y_{i,j}$ for $EF~~ 2p\sigma 2p\sigma^2$ state, outer well. $E_{max}=118376.981$.}
\begin{tabular}{c|ccccccc} 
\hline
$i \backslash j$ & 0&1&2&3&4&5&6\\ 
\hline 
0 & 98606.2 & 5.74637 & 1.37059 & -0.17094 & 0.00474712 & 0.000128489 & -4.8543e-6\\
1  & 1454.02 & -2.7411 & -0.180194 & 0.022727 & -0.00104196 & 1.65094e-5\\         
2  & -80.3398 & 1.01425 & 0.00144103 & -0.000222756 & 5.37232e-7\\               
3  & 1.77242 & -0.101008 & 0.000235334 & 5.67045e-6\\                          
4  & 0.191269 & 0.00388008 & -9.85281e-6\\                                   
5  & -0.0119227 & -5.07168e-5\\                                            
6  & 0.000188356\\
\hline 
\end{tabular}
\label{tab:dunham-EFouter}
\end{table*}}

\onltab{
\begin{table*}[h]
\scriptsize
\caption{$Y_{i,j}$ for $B'~~ 3p\sigma$ state. $E_{max}=133610.273$}
\begin{tabular}{c|cccccccc} 
\hline
$i \backslash j$ & 0&1&2&3&4&5&6&7\\ 
\hline 
0  & 109640.0 & 28.872 & 0.836305 & -0.0136974 & 4.01261e-5 & 1.23323e-6 & -1.6454e-8 & 8.27886e-11\\
1  & 1427.0 & -15.3274 & -2.85783 & 0.0525351 & -0.000479745 & 2.55235e-6 & -7.52893e-9 & 1.07135e-11\\           
2  & 689.282 & 27.9589 & 2.41683 & -0.0353017 & 0.000231187 & -7.70384e-7 & 8.98726e-10\\                       
3  & -449.987 & -22.3981 & -0.880946 & 0.00944058 & -3.7779e-5 & 6.29088e-8\\                                 
4  & 136.699 & 8.40707 & 0.160582 & -0.00115863 & 2.0487e-6\\                                               
5  & -21.732 & -1.6006 & -0.0142388 & 5.45786e-5\\                                                        
6  & 1.61573 & 0.149031 & 0.000481531\\                                                                 
7  & -0.0356702 & -0.00537226\\                                                                       
8  & -0.00077895\\
\hline
\hline
$i \backslash j$ &8\\
\hline
0& -1.55865e-13\\
\hline 
\end{tabular}
\label{tab:dunham-3psigma}
\end{table*}}

\onltab{
\begin{table*}[h]
\scriptsize
\caption{$Y_{i,j}$ for $D~~ 3p\pi$ state.  $E_{max}=133610.273$.}
\begin{tabular}{c|cccccccc} 
\hline
$i \backslash j$ & 0&1&2&3&4&5&6&7\\ 
\hline 
0  & 111713.0 & 30.0272 & -0.0142088 & -2.07092e-5 & -5.35965e-8 & 1.77885e-9 & -8.13272e-12 & 1.15704e-14\\
1  & 2354.63 & -0.771873 & -0.0151207 & 0.000144998 & -7.04225e-7 & 1.31303e-9 & 1.27466e-12 & -5.18074e-15\\           
2  & -68.7361 & -0.431929 & 0.00809662 & -6.33383e-5 & 2.90083e-7 & -7.21926e-10 & 6.91788e-13\\                      
3  & 2.91624 & 0.0927337 & -0.00152849 & 8.35883e-6 & -2.25164e-8 & 2.73782e-11\\                                   
4  & -0.743664 & -0.0073304 & 0.000130395 & -4.7786e-7 & 5.34873e-10\\                                            
5  & 0.103972 & 8.18834e-5 & -4.89806e-6 & 1.01828e-8\\                                                         
6  & -0.00749572 & 1.39201e-5 & 6.07114e-8\\                                                                  
7  & 0.000262281 & -4.49286e-7\\                                                                            
8  & -3.56422e-6\\
\hline
\hline
$i \backslash j$ &8\\
\hline
0 & 4.25363e-20\\
\hline 
\end{tabular}
\label{tab:dunham-3ppi}
\end{table*}}

\onltab{
\begin{table*}[h]
\scriptsize
\caption{$Y_{i,j}$ for $GK~~ 3d\sigma$ state, inner well. $E_{max}=116900.0$.}
\begin{tabular}{c|ccccc} 
\hline
$i \backslash j$ & 0&1&2&3&4\\ 
\hline 
0  & 109217.0 & 23.8085 & -0.253881 & -0.00473545 & 1.98913e-5\\
1  & 970.614 & -7.78821 & 0.412418 & 4.68762e-5\\             
2  & 24.2579 & -1.06924 & -0.0590925\\                      
3  & 1.64268 & 0.369018\\                                 
4  & -0.465966\\
\hline 
\end{tabular}
\label{tab:dunham-GK-inner}
\end{table*}}

\onltab{
\begin{table*}[h]
\scriptsize
\caption{$Y_{i,j}$ for $GK~~ 3d\sigma$ state, outer well.}
\begin{tabular}{c|ccccc} 
\hline
$i \backslash j$ & 0&1&2&3&4\\ 
\hline 
0  & 110177.0 & 14.5226 & 0.387782 & -0.0137014 & 0.000137787\\
1  & 1009.41 & -7.16505 & -0.0941763 & 0.00148123\\          
2  & -25.2145 & 2.75273 & 0.00636093\\                     
3  & -1.88729 & -0.245496\\                              
4  & 0.489779\\
\hline 
\end{tabular}
\label{tab:dunham-GK-outer}
\end{table*}}

\onltab{
\begin{table*}[h]
\scriptsize
\caption{$Y_{i,j}$ for $H\overline{H}~~ 3s\sigma$ state, inner well. $E_{max}=118000.0$.}
\begin{tabular}{c|ccccc} 
\hline
$i \backslash j$ & 0&1&2&3&4\\ 
\hline 
0  & 111909.0 & 33.4756 & -0.0628269 & -0.00674187 & 8.95921e-5\\
1  & 2003.13 & -12.1054 & 0.479189 & 0.00472823\\              
2  & 199.147 & 2.47998 & -0.246944\\                         
3  & 1.52825 & 0.64834\\                                   
4  & -22.8763\\
\hline 
\end{tabular}
\label{tab:dunham-HH-inner}
\end{table*}}

\onltab{
\begin{table*}[h]
\scriptsize
\caption{$Y_{i,j}$ for $H\overline{H}~~ 3s\sigma$ state, outer well.}
\begin{tabular}{c|ccccccc} 
\hline
$i \backslash j$ & 0&1&2&3&4&5&6\\ 
\hline 
0  & 122617.0 & 2.85838 & -0.452462 & 0.0496827 & -0.00129124 & -3.65898e-5 & 1.30097e-6\\
1  & 600.772 & -2.03478 & 0.165172 & -0.0186535 & 0.000900787 & -1.33774e-5\\           
2  & -178.292 & 0.705169 & -0.0128 & 8.90067e-5 & -7.25588e-6\\                       
3  & 48.1293 & -0.0969926 & 0.00148014 & 1.46566e-5\\                               
4  & -5.82021 & 0.00514187 & -6.94023e-5\\                                        
5  & 0.319828 & -8.17859e-5\\                                                   
6  & -0.00657088\\ 
\hline 
\end{tabular}
\label{tab:dunham-HH-outer}
\end{table*}}

\onltab{
\begin{table*}[h]
\scriptsize
\caption{$Y_{i,j}$ for $I~~ 3d\pi$ state.}
\begin{tabular}{c|ccccc} 
\hline
$i \backslash j$ & 0&1&2&3&4\\ 
\hline 
0  & 110177.0 & 14.5226 & 0.387782 & -0.0137014 & 0.000137787\\
1  & 1009.41 & -7.16505 & -0.0941763 & 0.00148123\\          
2  & -25.2145 & 2.75273 & 0.00636093\\                     
3  & -1.88729 & -0.245496\\                              
4  & 0.489779\\
\hline 
\end{tabular}
\label{tab:dunham-I}
\end{table*}}

\onltab{
\begin{table*}[h]
\scriptsize
\caption{$Y_{i,j}$ for $J~~ 3d\delta$ state.}
\begin{tabular}{c|cccc} 
\hline
$i \backslash j$ & 0&1&2&3\\ 
\hline 
0  & 111260.0 & 41.8057 & 0.0950125 & -0.00450042\\
1  & 2006.46 & -6.68957 & -0.0253833\\           
2  & 195.159 & 0.807222\\                      
3  & -56.9367\\ 
\hline 
\end{tabular}
\label{tab:dunham-J}
\end{table*}}

\onltab{
\begin{table*}[h]
\scriptsize
\caption{$Y_{i,j}$ for $B"Bbar~~ 4p\sigma$ state, inner well. $E_{max}=130265.0$.}
\begin{tabular}{c|ccccccc} 
\hline
$i \backslash j$ & 0&1&2&3&4&5&6\\ 
\hline 
0  & 115722.0 & 116.926 & -4.45735 & 0.188078 & -0.00440882 & 4.63197e-5 & -1.71816e-7\\
1  & 5867.53 & -52.4159 & -0.493037 & 0.0487225 & -0.000747071 & 3.25362e-6\\         
2  & -4015.06 & 44.4104 & -0.526875 & 0.00647721 & -2.37322e-5\\                    
3  & 1961.35 & -13.9427 & 0.00852245 & -9.78352e-5\\                              
4  & -482.848 & 2.58965 & 0.000454106\\                                         
5  & 56.5459 & -0.17409\\                                                     
6  & -2.50753\\ 
\hline 
\end{tabular}
\label{tab:dunham-BBbar_inner}
\end{table*}}

\onltab{
\begin{table*}[h]
\scriptsize
\caption{$Y_{i,j}$ for $B"Bbar~~ 4p\sigma$ state, outer well. $E_{max}=138941.911$.}
\begin{tabular}{c|ccccccc} 
\hline
$i \backslash j$ & 0&1&2&3&4&5&6\\ 
\hline 
0  & 122624.0 & 0.95644 & 0.000305579 & -3.2579e-6 & -4.94167e-8 & 1.00874e-9 & -4.52608e-12\\
1  & 360.513 & 0.00827397 & -6.93853e-5 & 1.30952e-6 & -1.07964e-8 & 3.57405e-11\\          
2  & -4.41936 & -0.00140404 & 1.00237e-6 & -2.28296e-8 & 5.20907e-11\\                    
3  & 0.0202615 & 6.54562e-5 & 5.32386e-8 & 2.54713e-10\\                                
4  & 0.000542329 & -1.81073e-6 & -1.7598e-9\\                                         
5  & -1.29422e-5 & 2.24736e-8\\                                                     
6  & 8.5522e-8\\ 
\hline 
\end{tabular}
\label{tab:dunham-BBar-outer}
\end{table*}}

\onltab{
\begin{table*}[h]
\scriptsize
\caption{$Y_{i,j}$ for $D'~~ 4p\pi$ state. $E_{max}=138941.911$.}
\begin{tabular}{c|ccccccc} 
\hline
$i \backslash j$ & 0&1&2&3&4&5&6\\ 
\hline 
0  & 116682.0 & 30.2145 & -0.0194539 & -0.000248291 & 8.12785e-6 & -8.99861e-8 & 3.37172e-10\\
1  & 2343.11 & -1.70002 & 0.00414816 & -8.34473e-5 & 9.02456e-7 & -3.67685e-9\\             
2  & -71.7393 & 0.0665521 & 0.000201895 & -1.68829e-6 & 1.27848e-8\\                      
3  & 2.55445 & -0.0143264 & -3.77337e-5 & 1.03224e-8\\                                  
4  & -0.348338 & 0.00198268 & 2.02175e-6\\                                            
5  & 0.0287842 & -0.000102487\\                                                     
6  & -0.000981186\\ 
\hline 
\end{tabular}
\label{tab:dunham-4ppi}
\end{table*}}

\onltab{
\begin{table*}[h]
\scriptsize
\caption{$Y_{i,j}$ for $5p\sigma$ state. $E_{max}=143570.0$.}
\begin{tabular}{c|ccccccc} 
\hline
$i \backslash j$ & 0&1&2&3&4&5&6\\ 
\hline 
0  & 118948.0 & 12.553 & -0.458717 & -0.0753537 & -0.00159359 & 9.37272e-5 & 9.54153e-6\\
1  & 1192.64 & 36.5609 & 3.12137 & 0.00958014 & -0.00304086 & -0.00015569\\            
2  & 956.798 & -40.5327 & -0.987989 & 0.015968 & 0.00140718\\                        
3  & -419.015 & 13.7138 & 0.0387677 & -0.00421793\\                                
4  & 83.7293 & -1.72833 & 0.00539521\\                                           
5  & -7.99491 & 0.072457\\                                                     
6  & 0.288787\\ 
\hline 
\end{tabular}
\label{tab:dunham-5psigma}
\end{table*}}

\onltab{
\begin{table*}[h]
\scriptsize
\caption{$Y_{i,j}$ for $5p\pi$ state. $E_{max}=143570.0$.}
\begin{tabular}{c|ccccccc} 
\hline
$i \backslash j$ & 0&1&2&3&4&5&6\\ 
\hline 
0  & 118995.0 & 36.0716 & -0.319741 & -0.014872 & -0.000238203 & 1.99987e-5 & 2.23272e-6\\
1  & 2415.58 & 2.15448 & 0.505586 & 0.0148412 & -0.000399269 & -6.86912e-5\\            
2  & -147.992 & -7.47147 & -0.160739 & 0.000242203 & 0.000546545\\                    
3  & 36.0906 & 2.80197 & 0.00598473 & -0.00176365\\                                 
4  & -7.57518 & -0.384299 & 0.0030262\\                                           
5  & 0.760411 & 0.016345\\                                                      
6  & -0.0284743\\ 
\hline 
\end{tabular}
\label{tab:dunham-5ppi}
\end{table*}}

\onltab{
\begin{table*}[h]
\scriptsize
\caption{$Y_{i,j}$ for $6p\sigma$ state. $E_{max}=148240.0$.}
\begin{tabular}{c|ccccccc} 
\hline
$i \backslash j$ & 0&1&2&3&4&5&6\\ 
\hline 
0  & 120709.0 & 27.2762 & -0.142997 & 0.00648973 & 0.000194723 & -2.7412e-6 & -6.47774e-7\\
1  & 61.5226 & -15.6536 & 0.324584 & 0.00741685 & -9.39748e-5 & -1.86748e-5\\            
2  & 2379.78 & 7.34398 & -0.0986585 & 0.00318703 & -2.58488e-5\\                       
3  & -1170.3 & -1.19031 & -0.0159069 & 1.37776e-5\\                                  
4  & 269.026 & 0.074286 & 0.00197083\\                                             
5  & -29.2007 & -0.000671958\\                                                   
6  & 1.20284\\ 
\hline 
\end{tabular}
\label{tab:dunham-6psigma}
\end{table*}}

\onltab{
\begin{table*}[h]
\scriptsize
\caption{$Y_{i,j}$ for $6p\pi$ state. $E_{max}=148240.0$.}
\begin{tabular}{c|ccccccc} 
\hline
$i \backslash j$ & 0&1&2&3&4&5&6\\ 
\hline 
0  & 120320.0 & 58.0291 & -0.686311 & -0.0354304 & -0.00063191 & 3.08385e-5 & 3.85789e-6\\
1  & 2264.49 & -53.1853 & 1.41237 & 0.0776417 & -1.83826e-5 & -0.000191495\\            
2  & 15.0754 & 28.5592 & -0.725983 & -0.027318 & 0.00157845\\                         
3  & -37.0221 & -6.48257 & 0.162007 & -0.00265325\\                                 
4  & 7.77546 & 0.602542 & -0.00483593\\                                           
5  & -0.725696 & -0.0217911\\                                                   
6  & 0.0252225\\
\hline 
\end{tabular}
\label{tab:dunham-6ppi}
\end{table*}}

\onltab{
\begin{table*}[h]
\scriptsize
\caption{$Y_{i,j}$ for $7p\sigma$ state. $E_{max}=150000.0$.}
\begin{tabular}{c|ccccccc} 
\hline
$i \backslash j$ & 0&1&2&3&4&5&6\\ 
\hline 
0  & 120984.0 & 14.4145 & -0.206167 & 0.0189087 & 0.000577862 & 9.43236e-7 & -1.05878e-6\\
1  & 2253.23 & 9.04999 & 0.299251 & 0.00127009 & -0.000479448 & -4.18786e-5\\           
2  & -32.7638 & -6.58075 & -0.230493 & 0.00195881 & 0.000781566\\                     
3  & -8.66359 & 2.25858 & 0.021996 & -0.00369885\\                                  
4  & 1.34773 & -0.280184 & 0.00674419\\                                           
5  & -0.16511 & 0.00597372\\                                                    
6  & 0.0113496\\ 
\hline 
\end{tabular}
\label{tab:dunham-7psigma}
\end{table*}}

\onltab{
\begin{table*}[h]
\scriptsize
\caption{$Y_{i,j}$ for $7p\pi$ state. $E_{max}=150000.0$.}
\begin{tabular}{c|ccccccc} 
\hline
$i \backslash j$ & 0&1&2&3&4&5&6\\ 
\hline 
0  & 121025.0 & 54.6852 & -0.657303 & -0.0278393 & -0.000517054 & 1.93446e-5 & 2.61651e-6\\
1  & 2526.84 & -33.3826 & 0.975964 & 0.0535573 & 0.000185217 & -0.000113024\\            
2  & -237.926 & 18.9794 & -0.678591 & -0.0232719 & 0.00118664\\                        
3  & 63.6492 & -4.60154 & 0.1826 & -0.00237026\\                                     
4  & -11.6711 & 0.422664 & -0.00727658\\                                           
5  & 1.08902 & -0.0152207\\                                                      
6  & -0.040398\\ 
\hline 
\end{tabular}
\label{tab:dunham-7ppi}
\end{table*}}

\begin{figure}[h]
\centering
\includegraphics[width=\hsize]{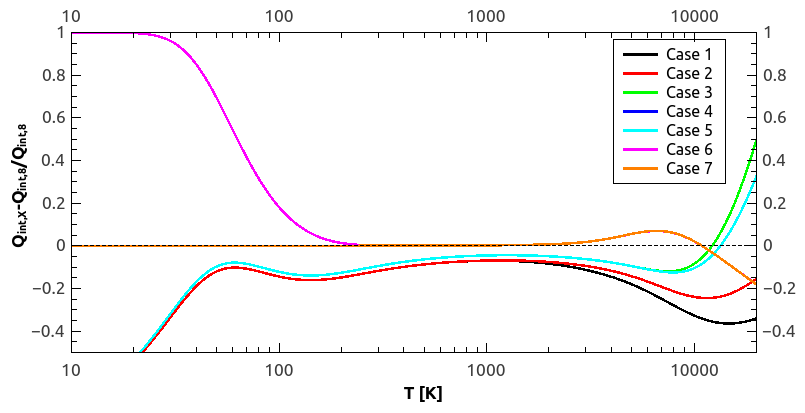}
\caption[$\rm H_2$ PF]{Comparison of different approaches for calculating $Q_{int}$ for the $\rm H_2$ molecule. Case 1 is identical to case 2 below 1500 K; case 4 is completely identical to case 5 for $\rm H_2$. All of the methods (i.e. cases 1-7) diverge substantially from our most accurate computation (case 8) for low as well as high temperatures.}
\label{fig:H2pf}
\end{figure}

\begin{figure*}
   \centering
                {\includegraphics[width=17cm]{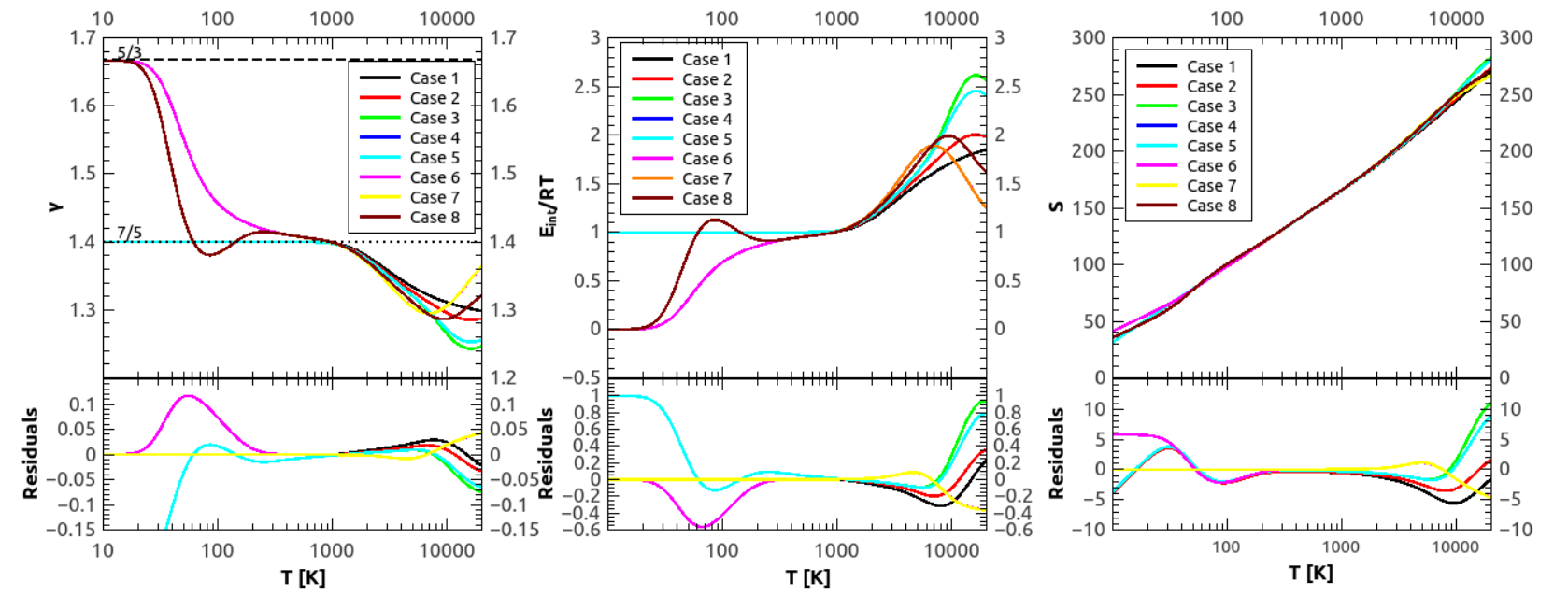}}
\caption{Adiabatic index (left panel), normalised internal energy (middle panel), and entropy (right panel) for the $\rm H_2$ molecule. In the left panel we also show the constants $\gamma=5/3$ and $\gamma=7/5$ . Residuals are for cases 1-7 with respect to case 8. In the left panel cases 1-5 are equal to $\gamma = 7/5$ below 1000 K, case 6 is identical to case 7 after 300 K, and case 7 is identical to case 8 until 6000 K; in the middle panel cases 1 to 5 are identical until 1000 K, and case 7 is almost identical to case 8 until 2000 K.}
\label{fig:H2cases-thermo}
\end{figure*}

\subsection{Comparison of cases 1 to 8}
First, we consider only the equilibrium $\rm H_2$. The comparison
of cases 1-7 to case 8 is shown in Fig. \eqref{fig:H2pf}. The main differences between all cases come from whether or not rigid-rotor approximation is used. The simplest case, case 1, underestimates $Q_{int}$ throughout the complete temperature range. The underestimation in the high-temperature range is easily explained by the equidistant separation (or linear increase with increasing $J$) between eigenstates (Eqs. \eqref{eq:vib-en-sep} and \eqref{eq:rot-en-sep}, respectively): in HOA/RRA the  eigenstate energies are higher for a given set of $\rm v$, $J$ values than when less strict approximations are used (Eq. \eqref{eq:vibration_energy_separations}). This effect is particularly strong for $H_2$, since the rotational distortion constant $D_e$ (see Eq. \eqref{eq:nonrig_rot}) is so large ($D_{e,\rm {H_2}} = 0.0471$ for the $X ^1\Sigma_g$ state, whereas it is only $D_{e, \rm {CO}}=6.1215\times 10^{-6}$ in the case of CO). The higher energy levels are therefore less populated in the HOA/RRA than in a more realistic computation, and these levels play a relatively larger role the higher the temperature. At the lowest temperatures the few lowest eigenstates dominate the value of $Q_{int}$, and these levels have slightly too high energies in HOA/RRA and therefore also cause $Q_{int}$ from HOA/RRA to
be substantially underestimated for low temperatures. Numerical differences are small, but the percentage difference is substantial: between 80 K and 1200 K the difference is 10 to 20 percent. Anharmonicity (difference between cases 1 and 2) for $\rm H_2$ is only relevant above 1000 K. Interaction between rotation and vibration gives only 2 percent difference (between cases 2 and 3) below 2000 K and becomes substantial thereafter. Since $D_0=36118.0696$ cm$^{-1}$, and $T_e = 91700$  cm$^{-1}$ for the $B ^1\Sigma_u^+$ state, for $\rm H_2$ it makes no difference whether excited electronic states are taken into account or not for low temperatures or if arbitrary truncation (e.g. 40000 cm$^{-1}$) is below the first excited state. On the other hand, summing above the dissociation energy, at temperatures above 10000 K gives larger $Q_{int}$ than it should be, assuming only the ground electronic state is available. On the other hand, if the RRA is used, $Q_{int}$ is still somewhat too low even if erroneously, it is summed to infinity. Case 6 shows how large the difference is when nuclear spin-split degeneracy is omitted from the full calculations and $\sigma=1/2$ is used. For temperatures below 200 K, case 6 overestimates the partition function by up to a factor of 2 compared to case 8. For T between 200 K and 2000 K, cases 6 and 7 are indistinguishable from case 8. Above 2000 K even case 7 (and 6) gives different results. The latter is due to poor predictions of the higher rotational energy states at higher vibrational levels.\\
Naturally, using different approaches (cases 1-8) leads to different estimates of thermodynamic quantities. In Fig. \ref{fig:H2cases-thermo} the adiabatic index $\gamma$, the normalised internal energy ($E_{int}/RT),$ and the entropy $S$ are shown for all the cases. Constant adiabatic indexes (5/3 or 7/5) clearly completely misrepresent reality for a wide range of temperatures when local thermal equilibrium (LTE) is assumed.  When considering the simplest cases (case 1 to 5), they follow the 7/5 simplification well until 1000 K, but are far from the most realistic estimate (case 8). Case 6 gives results almost identical to case 7 above 300 K. Case 7 is indistinguishable from case 8 up until 2000 K. The same trends can be seen for internal energy. From a first glance, the entropy seems to be very similar in all the cases, but this is because the dominating factor comes from $Q_{tr}$. Keeping that in mind, variability throughout all the temperature range is substantial. Entropy in case 7 is accurate until 3000 K.
\begin{figure}[h]
\centering
\includegraphics[width=\hsize]{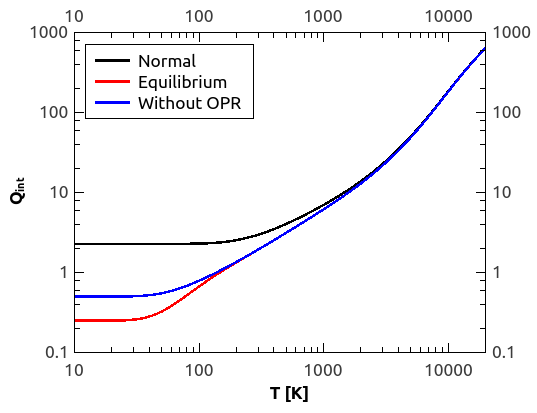}
\caption[$\rm H_2$ PF]{$Q_{int}$ for equilibrium, normal, and one (i.e. neglecting OPR) hydrogen.}
\label{fig:Q-normal-eq-noOP}
\end{figure}

\begin{figure}[h]
\centering
\includegraphics[width=\hsize]{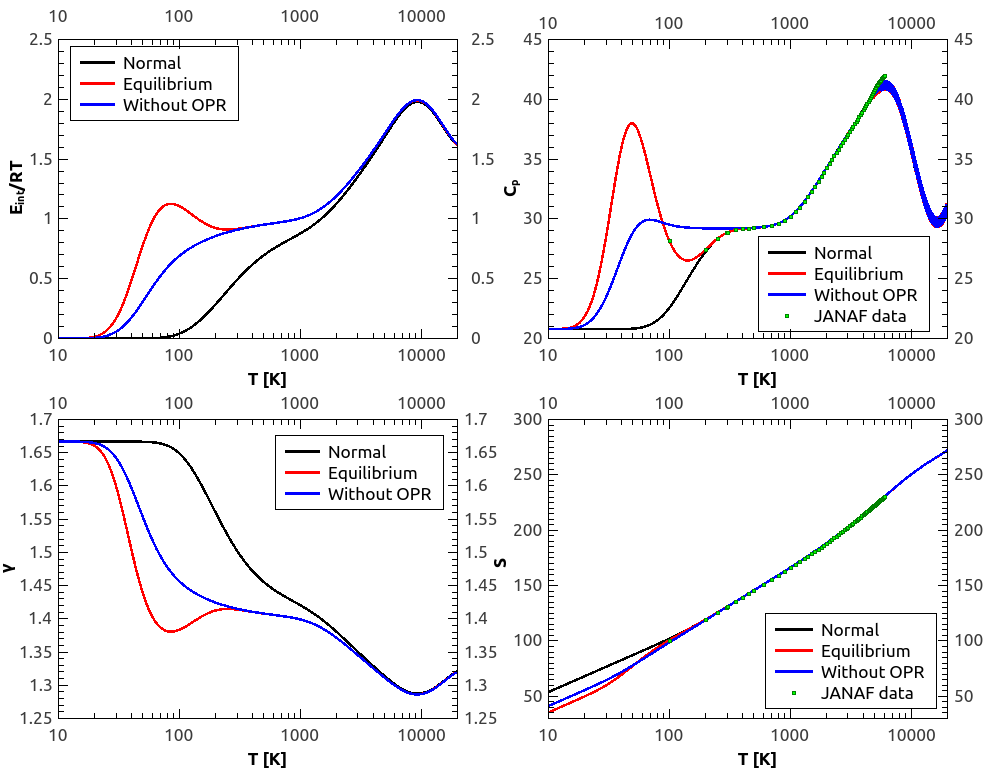}
\caption[$\rm H_2$ PF]{Normalised internal energy, $C_p$, adiabatic index and entropy, calculated for equilibrium, normal, and with neglected OPR  hydrogen. For $C_p$ and entropy JANAF data are also shown.}
\label{fig:thermo-normal-eq-noOP}
\end{figure}

        Now we compare equilibrium and normal hydrogen. Equilibrium, normal, and one, neglecting OPR, partition functions are shown in Fig. \eqref{fig:Q-normal-eq-noOP}. At low temperature, the three models show large differences, while in the high temperature range (T $>$ 2000 K) the results coalesce. The same trends can be seen for the calculated thermodynamic quantities (Fig. \eqref{fig:thermo-normal-eq-noOP}). Both internal energy and specific heat at constant pressure are lower for normal hydrogen, whereas equilibrium hydrogen has higher values. The adiabatic index is dramatically different for both cases. Normal hydrogen does not have a depression at low temperatures. This depression can lead to accelerated  gravitational instability and collapse in molecular clouds to form pre-stellar cores and protoplanetary cores in protoplanetary disks. Large differences in entropy are only present at temperatures below 300 K. Neglecting OPR would lead to intermediate values between normal and equilibrium hydrogen.

\subsubsection{Other studies.} There are a number of commonly used studies of $Q_{int}$. Here we give a brief summary of how other studies were performed, which simplifications they used, which molecular constants were adopted, and for which temperature range they were recommended. We stress that although it is commonly done, data must not be extrapolated beyond the recommended values
under any circumstances. The resulting partition functions as well as their derivatives often diverge from the true values remarkably fast outside the range of the data they are based on. In our present study we took care to validate our data and formulas, such that they can be used in a wider temperature range than thos of any previous study. Our data and formulas can be safely applied in the full temperature range from 1 K to 20000 K and easily extended beyond 20000 K in extreme cases, such as shocks, by use of the listed Dunham coefficients.

\paragraph{Tatum 1966 \cite[T1966]{Tatum1966}} presented partition functions and dissociation equilibrium constants for 14 diatomic molecules in the temperature range $T = 1000 - 8000$ K with 100 K steps. Calculations were made using case 5. The upper energy cut-off was chosen to be at $40000$ cm$^{-1}$. Only the first few molecular constants were used ($\omega_e, \omega_ex_e, B_e, \alpha_e$).

\paragraph{Irwin 1981\cite[I1981]{Irwin1981}} presented partition function approximations for 344 atomic and molecular species, valid for the temperature range  $T = 1000 - 16000$ K. Data for molecular partition functions were taken from \cite{Tatum1966,McBride1963,JANAF}. \cite{Irwin1981} gave partition function data in the form of polynomials,
\begin{equation}
ln Q = \sum_{i=0}^5 a_i(ln T)^i,
\label{eq:irwin_poly}
\end{equation}
which is very convenient to use. For each species, the coefficients of Eq. \eqref{eq:irwin_poly} were found by the method of least-squares fitting. Some molecular data were linearly extrapolated before fitting the coefficients. For the extrapolated data the weight was reduced by a factor of $10^6$. Irwin stated that these least-squares fits have internally only small errors. However, a linear extrapolation to strongly non-linear data results in large errors even before fitting the polynomial, and the size of the errors due to this effect was not estimated.

\paragraph{Bohn and Wolf 1984 \cite[BW1984]{Bohn1984}} derived partition functions for $\rm H_2$ and $\rm CO$ that are valid for the temperature range $T = 1000 - 6000$ K. In principle, this paper aimed to show a new approximate way of calculating partition functions, specific heat $c_V$ , and internal energy $E_{int}$. Calculations were made using case 5 for "exact" partition functions and using assumptions of case 3 for approximations. Only the ground electronic state was included, and only the first few molecular constants were used ($\omega_e, \omega_ex_e, B_e, \alpha_e$).

\paragraph{Sauval and Tatum 1984 \cite[ST1984]{Sauval1984}} presented total internal partition functions for 300 diatomic molecules, 69 neutral atoms, and 19 positive ions. Molecular constants ($\omega_e, \omega_ex_e, B_e, \alpha_e$) were taken from \cite{Huber1979}. The partition function polynomial approximations are valid for the temperature range T $= 1000 - 9000$ K. The authors adopted the previously used equation, [T1966] (case 5) for all the species. A polynomial expression
\begin{equation}
log Q = \sum^4_{n=0} a_n (log \theta)^n
\label{eq:sauvtat1984}
\end{equation}
was used for both atoms and molecules. Here $\theta = 5040/ T$.

\paragraph{Rossi et al. 1985 \cite[R1985]{Rossi1985}} presented total internal partition functions for 53 molecular species in the temperature range T $= 1000 - 5000$ K. Molecular constants ($\omega_e, \omega_ex_e, B_e, \alpha_e$) were taken from \cite{Huber1979}. For diatomic molecules the authors followed Tatum's \cite{Tatum1966} paradigm, case 5. The authors claimed that $Q_{J}$ was evaluated with a non-rigid approximation \cite{Rossi1985}, but upon a simple inspection it is clear that they used rigid-rotor approximation. They did, however, allow interaction between vibration and rotation. An arbitrary truncation in the summation over the electronic states is at $40000$ cm$^{-1}$. For polynomial approximations the authors considered the "exact" specific heat, whose behaviour near the origin is more amenable to approximation schemes. The partition functions were then obtained by integration \cite{Rossi1985}. Their specific heat approximation is
\begin{equation}
\frac{C_{\rm v}}{k} = \sum^4_{m=0} a_mZ^m,
\end{equation}
where $Z = T/1000$ and $a_m$ are coefficients. The partition function they obtained is listed as the seven filling constants $a_0$ to $a_6$ to the polynomial:
\begin{equation}
ln Q = a_0 ln Z + \sum^4_{m=1} \frac{a_m}{m(m+1)}Z^m-\frac{a_5}{Z}+a_6.
\end{equation}

\paragraph{Kurucz 1985 \cite[K1985]{Kurucz1985}} commented on the situation regarding the partition functions of $\rm H_2$ and $\rm CO$. Approximate expressions for the molecular partition functions in previous studies were considered not rigorous enough because they did not include coefficients of sufficiently high order and did not keep proper track of the number of bound levels. K1985 used experimentally determined energy levels, when available, and supplemented them with fitted values, presumably extended to a higher cut-off energy as discussed in Sect. 2.6 and Fig.\,1 above. The author explicitly summed over the energies of the three lowest electronic states (data for $H_2$ were derived from \cite{Dabrowski1984}) and gave polynomial fits for the partition functions between 1000 K and 9000 K and also tabulated the exact results with steps of $100$ K.

\paragraph{Irwin1987 \cite[I1987]{Irwin1987}} Irwin (1987) presented refined total internal partition functions for $\rm H_2$ and $\rm CO$. The partition function polynomial approximations are valid for the temperature range T = 1000 - 9000 K. Estimated errors at 4000 K are 2\% for $\rm H_2$ and 0.004\% for CO. $Y_{i,j}$ for $\rm H_2$ were determined by using an equally weighted simultaneous least-squares fit of Dunham series using energy data obtained from \cite{Dabrowski1984}. This was done for the ground $\rm H_2$ electronic state only.\\
I1987 compared his results with those of \cite{Kurucz1985,Sauval1984} and found slightly higher values of $Q$, which he attributed to a combination of the number of included electronic states and a difference in the treatments of the highest rotational levels. Our results indicate that the main difference between the results of I1987\cite{Irwin1987} and K1985\cite{Kurucz1985} is the use of slightly too low energies in I1987 of the highest rotational levels, and we conclude that the value trend by K1985 is slightly closer to the values obtained by a full treatment (our case 8) than are those of I1987. A main difference between our method and those of I1987, K1985, and older works is, however, that our method is applicable in a higher temperature range and can treat ortho and para states separately. It can therefore
be used in a wider range of physical conditions, including shocks, non-equilibrium gasses, etc.

\paragraph{Pagano et al. 2009\cite[P2009]{Pagano2009}} presented internal partition functions and thermodynamic properties of high-temperature (50 - 50000 K) Jupiter-atmosphere species. The authors used case 7 to calculate the partition functions with more than the first few (e.g.  $\omega_e$ and $\omega_ex_e$, $B_e$ and $D_e$) molecular constants for the ground and first few electronically excited states. The calculations are completely self-consistent in terms of maximum rotational states for the individual vibrational levels, presumably such that $E(\rm v, J_{max}) \approx D_0$ for each $\rm v$.

\paragraph{Laraia et al. 2011 \cite[L2011]{Laraia2011}} presented total internal partition functions  for a number of molecules and their isotopes for the temperature range $70 - 3000$ K. The methods used in this study are based on \cite{Gamache2000,Fischer2003}. Rotational partition functions were determined using the formulae in  \cite{McDowell1988,McDowell1990}. Vibrational partition functions were calculated using the harmonic oscillator approximation of \cite{Herzberg1960}. All state-dependent and state-independent degeneracy factors were taken into account in this study. The $\rm H_2$ partition function was calculated as a direct sum using \textup{ab initio }energies. A four-point Lagrange interpolation was used for the temperature range with intervals of 25 K. Data are presented in an easily retrievable FORTRAN program \cite{Laraia2011}.

\paragraph{Colonna et al. 2012\cite[C2012]{Colonna2012}} gave a statistical thermodynamic description of $\rm H_2$ molecules in normal ortho/para mixture. The authors determined the internal partition function for normal hydrogen on a rigorous basis, solving the existing ambiguity in the definition of those quantities directly related to partition functions rather than on its derivatives
\cite{Colonna2012}. The authors used case 7 with more molecular constants for the ground and first few electronically excited states than shown with equations \ref{eq:case7}. Internal partition function as well as thermodynamic properties for 5 - 10000 K are also given.


\begin{figure}[h]
\centering
\includegraphics[width=\hsize]{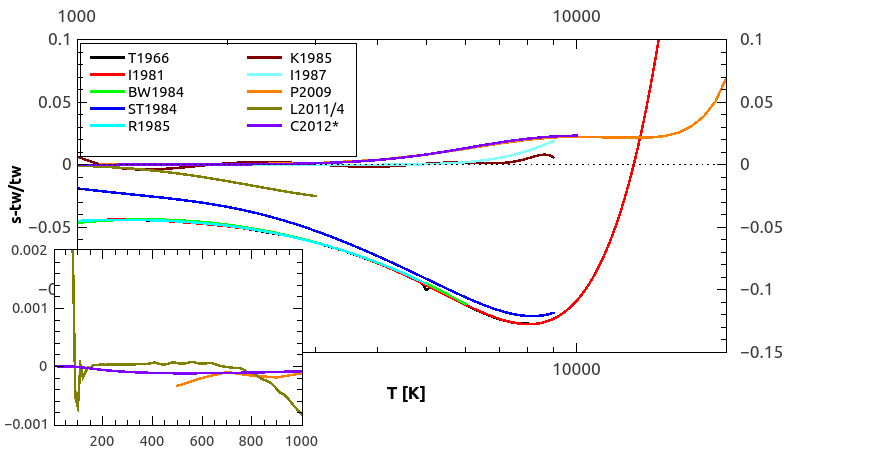}
\caption[$\rm H_2$ PF]{Normalised $Q_{int}$ from other studies in their respective temperature ranges. Normalisation is made with respect to the results of this work, ([study - this work]/this work). C2012 is normalised to normal hydrogen, all other studies to equilibrium hydrogen. The inset shows the low-temperature range.}
\label{fig:study-comparison}
\end{figure}

\paragraph{Comparison.}Figure \ref{fig:study-comparison}  shows the normalised value of $Q_{int}$ for molecular hydrogen from the ten different studies we have described here, in their respective temperature ranges listed in the respective studies. The normalisation is made with respect to the results of this work (case 8). For C2012 normal hydrogen $Q_{int}$ is used, whereas equilibrium hydrogen $Q_{int}$ is used for all the other studies. The main frame of Fig. \ref{fig:study-comparison} is for the 1000-20000K temperature range. The inset shows the low-temperature (10 - 1000 K) range. In the low-temperature range our results clearly differ only marginally from the three studies (P2009, L2011, C2012) that listed values of $Q_{int}$ for low temperatures (the percentage difference is smaller than 0.1 percent in most of the temperature range). Since L2011 did not normalise the spin-split degeneracy, their results had to be divided by a factor of 4 before the comparison, shown in Fig. \ref{fig:study-comparison}. In the high-temperature range,  all studies using cases 1-5  have large errors (also seen from our case comparison in Fig. \ref{fig:H2pf}). Since K1985 explicitly derived his partition functions from experimental energy data, his results are in very good agreement with our results (from case 8) for his given temperature range (1000 – 9000 K). I1987 is also in great agreement with both K1985 and our results for the same temperature range. Using the values of $Y_{ij}$ determined by I1987 gives results that perfectly agree with our work, as expected. At the higher end of the I1987 recommended temperature range, however, the polynomial approximation starts to have larger errors (the estimated 2 percent). At the highest temperature range, P2009 and C2012 also start to show considerable errors. These two studies are the only ones based on summation of energy levels, calculated based on molecular constants, and the comparison of our work with P2009 and C2012 therefore illustrates approximately how large errors in $Q_{int}$ are obtained from methods based on molecular constants. The reason is, as mentioned before, that molecular constants are commonly fitted to the bottom of the potential well and poorly represent higher vibrational and rotational states energies. Finally, the I1981 results follow the T1966 results perfectly until 8000 K, which is the latter's high temperature limit. Then, since I1981 extrapolated data, the corresponding $Q_{int}$ begin to show exponentially larger errors and is completely unreliable, which illustrates both how poor the HO + RR approximation represents reality (T1966 and I1981), and, in particular, \textit{\textup{how unreliable it can be to extrapolate beyond the calculated or measured data}} (as was done in the work of I1981), as seen clearly in Fig \ref{fig:study-comparison}.

\begin{figure}
\centering
\includegraphics[width =\hsize]{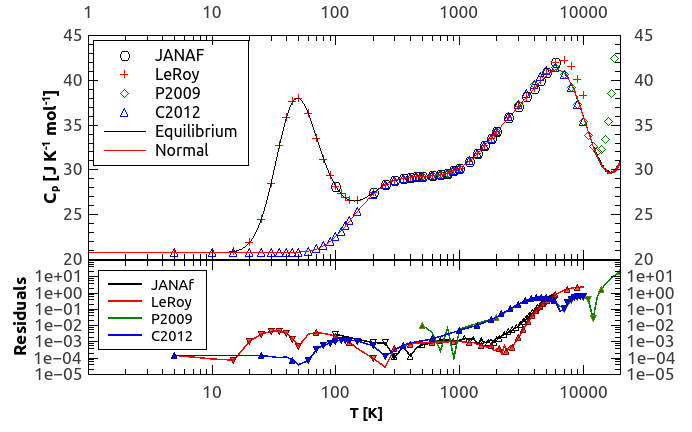}
\caption[$\rm H_2$ PF]{Calculated $C_p$ comparison to other studies. The top panel shows resulting $C_p$ values, the bottom panel residuals (other - ours) on a logarithmic scale. Facing down
triangles indicate that other results have lower values than ours, facing up triangles indicate that the values are higher.}
\label{fig:thermo-comparison-cp}
\end{figure}

\begin{figure}[ht]
\centering
\includegraphics[width =\hsize]{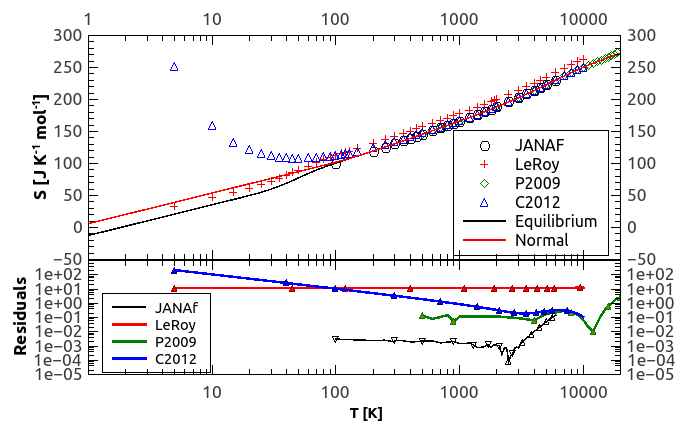}
\caption[$\rm H_2$ PF]{Same as in Fig. \ref{fig:thermo-comparison-cp}, but for $S$.}
\label{fig:thermo-comparison-S}
\end{figure}

In Fig. \ref{fig:thermo-comparison-cp} our results for $\rm H_2$ specific heat at constant pressure are compared to \cite[JANAF]{JANAF}, \cite[LeRoy]{LeRoy1990}, \cite[P2009]{Pagano2009} and \cite[C2012]{Colonna2012}. The latter is compared to our results for normal hydrogen, and the others are compared to our equilibrium hydrogen calculations. All calculations agree very well until 10000 K.  Figure \ref{fig:thermo-comparison-S} depicts our results for entropy calculations for $\rm H_2$. Once again, the agreement is very good (except with C2012 below 100 K), especially between our results and the JANAF data. However, there is a systematic offset between LeRoy and our results (and those of JANAF) of 11.52 J K$^{-1}$ mol$^{-1}$. This might be due to slightly different STP values, used by \cite{LeRoy1990}, resulting in slightly different $Q_{tr}$. We note that the JANAF data represent equilibrium $H_2$, while we computed equilibrium as well as normal and ortho and para $H_2$ (and span a much wider temperature range), which makes our data more generally applicable.

\subsection{Polynomial fits}
\begin{figure*}[ht]
\centering
\includegraphics[width =\hsize]{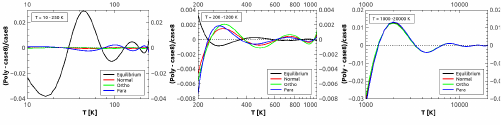}
\caption[$\rm H_2$ poly fits]{Accuracies of polynomial fits for the partition functions of $H_2$.}
\label{fig:our_poly_fits}
\end{figure*}

Following the common practice, we present polynomial fits to our results. All partition function data were fit to a fifth-order polynomial of the form
\begin{equation}
Q_{int}(T) = \sum_{i=0}^5 a_iT^i.
\end{equation}
The data for the partition function were spliced into three intervals to reach better accuracy. Polynomial constants are given in Table \ref{tab:our_Q_poly}, and the fitting accuracy is shown in Fig. \ref{fig:our_poly_fits}. Combining the three intervals, we reach
an error smaller than 2 percent for all the temperature range. The decision for using lookup tables with significantly larger accuracy or the faster polynomial fits with increased errors
needs to be made on an individual basis.

\begin{table*}[h]
\scriptsize
\caption{Polynomial fit constants for partition functions of equilibrium, normal, and  ortho and para $H_2$}
\centering
Low temperature (T = 10 - 250 K)
\begin{tabular}{c|ccccc|r} 
\hline
Flavor & a$_0$ & a$_1$ & a$_2$ & a$_3$ & a$_4$ & RMS\\ 
\hline 
Equilibrium & 2.673071615415136e-01 & -3.495586051757211e-03 & 1.227901619954258e-04 & -5.776440695273789e-07 & 9.251224490175610e-10& 0.00516\\
Normal & 2.277668085144430 & 3.115475456535298e-04 & -1.199204285095701e-05 & 1.348313176588763e-07 & -2.493525433376579e-10& 0.00091\\
Ortho & 2.996425936476927 & 3.807856081411796e-04 & -8.659737634951350e-06 & 5.779158662343921e-08 & -5.407749218414769e-11& 0.00107\\
Para & 9.994339093261912e-01 & 3.135419941617079e-04 & -1.599408560500201e-05 & 2.032015813340736e-07 & -4.090066248229891e-10 & 0.00186 \\ 
\hline 
\hline
\end{tabular}
Medium temperature (T = 200 - 1100 K)
\begin{tabular}{c|ccccc|r} 
\hline
Flavor & a$_0$ & a$_1$ & a$_2$ & a$_3$ & a$_4$ & RMS\\ 
\hline 
Equilibrium & 1.410033600294133e-01 & 6.085738724141971e-03 & -4.096994909866605e-07 & 4.220221708082499e-10 & -8.790594164685680e-14 & 0.0007\\
Normal & 1.919800108599199 & 1.746213811668277e-03 & 7.693197369847348e-06 & -6.504303232753276e-09 & 2.132168610356154e-12 & 0.00247\\
Ortho & 2.891845114721733 & -1.243647561012797e-03 & 1.312997140340565e-05 & -1.105110818469195e-08 & 3.562652399419052e-12 & 0.00402\\
Para & 3.312881928317837e-01 & 4.871438707876421e-03 & 2.339907793941012e-06 & -2.214128676054798e-09 & 8.210809799374644e-13 & 0.00164 \\ 
\hline 
\hline
\end{tabular}

High temperature (T = 1000 - 20000 K)
\begin{tabular}{c|ccccc|r} 
\hline
Flavor & a$_0$ & a$_1$ & a$_2$ & a$_3$ & a$_4$ & RMS\\ 
\hline 
Equilibrium & -9.661842638994980e-01 & 7.302127874247883e-03 & -6.760893004505151e-07 & 3.128741080316710e-10 & -1.645206030945910e-14 & 0.099 \\
Normal & -1.142827289681385e-01 & 7.245357485089481e-03 & -6.404547049175155e-07 & 3.110974834335349e-10 & -1.642844436813594e-14 &  0.10151\\
Ortho & 2.231696541394133e-01 & 7.190413006059544e-03 & -6.192076726407582e-07 & 3.096074912455146e-10 & -1.640036910200170e-14 &  0.10242\\
Para & -1.019053774741465e+00 & 7.360074236377774e-03 & -6.935790834485644e-07 & 3.145274660505019e-10 & -1.646581413255403e-14 &  0.10004 \\ 
\hline 
\hline
& a$_5$\\
\hline
& 2.788597060472472e-19\\
& 2.790315674249059e-19\\
& 2.790310846593324e-19\\
& 2.782419680521890e-19\\
\end{tabular}
\label{tab:our_Q_poly}
\end{table*}

\section{Conclusions}
We have investigated the shortcomings of various simplifications
that are used to calculate $Q_{int}$ and applied our analyses to calculate the time-independent partition function of normal, ortho and para, and equilibrium molecular hydrogen. We also calculated $E_{int}/RT$, $H-H(0)$, $S$, $C_p$, $C_v$, $-[G-H(0)]/T,$ and $\gamma$. Our calculated values of thermodynamic properties for ortho and para, equilibrium, and normal $H_2$ are presented in Tables \ref{tab:result-thermo-Qnorm}, \ref{tab:result-thermo-Qeq}, \ref{tab:result-thermo-Qortho}, and \ref{tab:result-thermo-Qpara}, rounded to three significant digits. Full datasets in 1 K temperature steps\footnote{Data with smaller temperature steps or beyond the given temperature range can be requested personally from the corresponding author} can be retrieved online from \cite{mydata}. The partition functions and thermodynamic data presented in this work are more accurate than previously available data from the literature and cover a more complete temperature range than any previous study in the literature. Determined Dunham coefficients for a number of electronic states of $\rm H_2$ are also reported.\\
In future work we plan to use the method we described here to calculate partition functions and thermodynamic quantities for other astrophysically important molecular species as well.

\onltab{
\begin{table*}[h]
\caption{Thermodynamic properties of equilibrium $\rm H_2$.}
\begin{tabular}{ccccccccc} 
\hline
$T$ & $Q_{int}$ & $E_{int}/RT$ & $H-H(0)$ & $S$ & $C_p$ & $C_v$ & $-[G-H(0)]/T$ &$\gamma$\\ 

[K] & & [J] & [J mol$^{-1}$] & [J K$^{-1}$ mol$^{-1}$] & [J $K^{-1}$ mol$^{-1}$] & [J K$^{-1}$ mol$^{-1}$] & [J K$^{-1}$ mol$^{-1}$]  \\
\hline
5.0 & 0.25 & 0.0 & 103.931 & 21.095 & 20.786 & 0.0 & -40.026 & 1.667\\
10.0 & 0.25 & 0.0 & 207.862 & 35.503 & 20.787 & 0.001 & 64.026 & 1.667\\
15.0 & 0.25 & 0.001 & 311.94 & 43.942 & 20.898 & 0.112 & 222.473 & 1.666\\
20.0 & 0.25 & 0.015 & 418.257 & 50.053 & 21.864 & 1.078 & 416.507 & 1.66\\
25.0 & 0.252 & 0.066 & 533.473 & 55.184 & 24.518 & 3.732 & 638.256 & 1.638\\
30.0 & 0.258 & 0.169 & 665.765 & 59.996 & 28.537 & 7.751 & 884.693 & 1.599\\
35.0 & 0.267 & 0.315 & 819.085 & 64.715 & 32.706 & 11.92 & 1154.933 & 1.551\\
40.0 & 0.282 & 0.48 & 991.15 & 69.305 & 35.893 & 15.107 & 1448.489 & 1.505\\
45.0 & 0.301 & 0.642 & 1175.533 & 73.647 & 37.61 & 16.824 & 1764.426 & 1.467\\
50.0 & 0.324 & 0.783 & 1364.975 & 77.638 & 37.97 & 17.184 & 2101.227 & 1.438\\
60.0 & 0.382 & 0.984 & 1737.824 & 84.441 & 36.242 & 15.456 & 2829.782 & 1.403\\
70.0 & 0.448 & 1.085 & 2086.766 & 89.826 & 33.537 & 12.75 & 3619.013 & 1.387\\
80.0 & 0.519 & 1.123 & 2409.737 & 94.142 & 31.153 & 10.367 & 4456.451 & 1.381\\
90.0 & 0.593 & 1.124 & 2711.881 & 97.703 & 29.377 & 8.591 & 5333.052 & 1.381\\
100.0 & 0.667 & 1.107 & 2999.115 & 100.73 & 28.151 & 7.365 & 6242.444 & 1.384\\
110.0 & 0.74 & 1.082 & 3276.322 & 103.373 & 27.351 & 6.565 & 7180.087 & 1.387\\
120.0 & 0.813 & 1.055 & 3547.19 & 105.73 & 26.867 & 6.081 & 8142.661 & 1.391\\
130.0 & 0.883 & 1.029 & 3814.429 & 107.869 & 26.613 & 5.827 & 9127.671 & 1.395\\
140.0 & 0.952 & 1.005 & 4080.006 & 109.837 & 26.526 & 5.739 & 10133.185 & 1.399\\
150.0 & 1.02 & 0.984 & 4345.331 & 111.668 & 26.556 & 5.77 & 11157.669 & 1.403\\
200.0 & 1.341 & 0.923 & 5692.821 & 119.414 & 27.448 & 6.662 & 16527.172 & 1.413\\
150.0 & 1.02 & 0.984 & 4345.331 & 111.668 & 26.556 & 5.77 & 11157.669 & 1.403\\
298.15 & 1.931 & 0.916 & 8467.176 & 130.682 & 28.836 & 8.05 & 28016.705 & 1.414\\
300.0 & 1.942 & 0.916 & 8520.534 & 130.86 & 28.849 & 8.063 & 28243.25 & 1.414\\
350.0 & 2.238 & 0.926 & 9969.563 & 135.327 & 29.081 & 8.294 & 34484.941 & 1.412\\
400.0 & 2.534 & 0.936 & 11426.436 & 139.218 & 29.181 & 8.395 & 40934.958 & 1.411\\
450.0 & 2.831 & 0.944 & 12886.795 & 142.658 & 29.229 & 8.442 & 47567.798 & 1.409\\
500.0 & 3.128 & 0.952 & 14349.02 & 145.739 & 29.259 & 8.473 & 54363.344 & 1.408\\
600.0 & 3.725 & 0.963 & 17278.057 & 151.079 & 29.326 & 8.54 & 68380.83 & 1.406\\
700.0 & 4.324 & 0.973 & 20215.836 & 155.608 & 29.44 & 8.654 & 82889.4 & 1.404\\
800.0 & 4.928 & 0.983 & 23168.335 & 159.55 & 29.623 & 8.837 & 97820.014 & 1.403\\
900.0 & 5.536 & 0.994 & 26142.864 & 163.053 & 29.88 & 9.094 & 113121.903 & 1.401\\
1000.0 & 6.151 & 1.006 & 29146.545 & 166.217 & 30.204 & 9.418 & 128756.479 & 1.399\\
1100.0 & 6.774 & 1.019 & 32185.337 & 169.113 & 30.579 & 9.793 & 144693.583 & 1.397\\
1200.0 & 7.406 & 1.034 & 35263.602 & 171.792 & 30.991 & 10.204 & 160909.037 & 1.395\\
1300.0 & 8.051 & 1.051 & 38384.094 & 174.289 & 31.421 & 10.635 & 177383.01 & 1.392\\
1400.0 & 8.709 & 1.069 & 41548.164 & 176.634 & 31.86 & 11.074 & 194098.888 & 1.389\\
1500.0 & 9.382 & 1.089 & 44756.054 & 178.847 & 32.297 & 11.511 & 211042.494 & 1.386\\
2000.0 & 13.015 & 1.193 & 61416.861 & 188.419 & 34.278 & 13.492 & 298792.538 & 1.371\\
2500.0 & 17.181 & 1.299 & 78962.143 & 196.243 & 35.838 & 15.052 & 390859.016 & 1.357\\
3000.0 & 21.964 & 1.397 & 97201.242 & 202.89 & 37.077 & 16.291 & 486526.148 & 1.345\\
3500.0 & 27.428 & 1.486 & 116007.0 & 208.686 & 38.121 & 17.335 & 585293.496 & 1.335\\
4000.0 & 33.632 & 1.568 & 135302.809 & 213.838 & 39.043 & 18.257 & 686790.762 & 1.326\\
4500.0 & 40.634 & 1.644 & 155031.8 & 218.484 & 39.852 & 19.066 & 790732.925 & 1.318\\
5000.0 & 48.492 & 1.713 & 175128.639 & 222.719 & 40.508 & 19.721 & 896892.151 & 1.311\\
6000.0 & 66.988 & 1.831 & 216042.905 & 230.176 & 41.156 & 20.37 & 1.115128382e6 & 1.3\\
7000.0 & 89.448 & 1.917 & 257096.393 & 236.506 & 40.779 & 19.993 & 1.340240895e6 & 1.293\\
8000.0 & 115.989 & 1.97 & 297322.673 & 241.879 & 39.553 & 18.767 & 1.571192049e6 & 1.288\\
9000.0 & 146.498 & 1.991 & 336046.785 & 246.442 & 37.836 & 17.049 & 1.807099542e6 & 1.286\\
10000.0 & 180.675 & 1.986 & 372961.362 & 250.333 & 35.979 & 15.193 & 2.047222929e6 & 1.287\\
11000.0 & 218.099 & 1.962 & 408052.021 & 253.679 & 34.211 & 13.425 & 2.29095485e6 & 1.289\\
12000.0 & 258.301 & 1.925 & 441490.549 & 256.589 & 32.676 & 11.89 & 2.537806546e6 & 1.292\\
13000.0 & 300.818 & 1.881 & 473549.464 & 259.156 & 31.484 & 10.698 & 2.787389971e6 & 1.296\\
14000.0 & 345.231 & 1.834 & 504545.723 & 261.453 & 30.59 & 9.804 & 3.039399948e6 & 1.3\\
15000.0 & 391.19 & 1.788 & 534806.189 & 263.541 & 29.996 & 9.21 & 3.293598228e6 & 1.304\\
16000.0 & 438.423 & 1.744 & 564646.464 & 265.467 & 29.722 & 8.936 & 3.549799968e6 & 1.308\\
17000.0 & 486.741 & 1.705 & 594357.51 & 267.269 & 29.744 & 8.958 & 3.80786255e6 & 1.312\\
18000.0 & 536.03 & 1.671 & 624196.974 & 268.974 & 29.964 & 9.178 & 4.067676383e6 & 1.315\\
19000.0 & 586.244 & 1.642 & 654383.656 & 270.606 & 30.448 & 9.662 & 4.329157371e6 & 1.318\\
20000.0 & 637.389 & 1.62 & 685094.498 & 272.181 & 39.3 & 18.514 & 4.59224074e6 & 1.321\\
\end{tabular}
\label{tab:result-thermo-Qeq}
\end{table*}}

\onltab{
\begin{table*}[h]
\caption{Thermodynamic properties of normal $\rm H_2$.}
\begin{tabular}{ccccccccc} 
\hline
$T$ & $Q_{int}$ & $E_{int}/RT$ & $H-H(0)$ & $S$ & $C_p$ & $C_v$ & $-[G-H(0)]/T$ &$\gamma$\\ 

[K] & & [J] & [J mol$^{-1}$] & [J K$^{-1}$ mol$^{-1}$] & [J $K^{-1}$ mol$^{-1}$] & [J K$^{-1}$ mol$^{-1}$] & [J K$^{-1}$ mol$^{-1}$]  \\
\hline
5.0 & 2.28 & 0.0 & 103.931 & 39.472 & 20.786 & 0.0 & 51.859 & 1.667\\
10.0 & 2.28 & 0.0 & 207.861 & 53.88 & 20.786 & 0.0 & 247.797 & 1.667\\
15.0 & 2.28 & 0.0 & 311.792 & 62.308 & 20.786 & -0.0 & 498.116 & 1.667\\
20.0 & 2.28 & 0.0 & 415.723 & 68.288 & 20.786 & 0.0 & 783.751 & 1.667\\
25.0 & 2.28 & 0.0 & 519.654 & 72.926 & 20.786 & 0.0 & 1095.646 & 1.667\\
30.0 & 2.28 & 0.0 & 623.585 & 76.716 & 20.786 & 0.0 & 1428.468 & 1.667\\
35.0 & 2.28 & 0.0 & 727.518 & 79.92 & 20.787 & 0.001 & 1778.693 & 1.667\\
40.0 & 2.28 & 0.0 & 831.461 & 82.696 & 20.791 & 0.005 & 2143.818 & 1.667\\
45.0 & 2.28 & 0.0 & 935.441 & 85.146 & 20.802 & 0.016 & 2521.971 & 1.667\\
50.0 & 2.28 & 0.0 & 1039.506 & 87.339 & 20.827 & 0.041 & 2911.705 & 1.666\\
60.0 & 2.28 & 0.002 & 1248.259 & 91.144 & 20.941 & 0.155 & 3721.536 & 1.666\\
70.0 & 2.281 & 0.006 & 1458.731 & 94.388 & 21.175 & 0.389 & 4566.442 & 1.664\\
80.0 & 2.284 & 0.014 & 1672.186 & 97.238 & 21.537 & 0.751 & 5441.706 & 1.661\\
90.0 & 2.29 & 0.026 & 1889.847 & 99.801 & 22.011 & 1.225 & 6343.964 & 1.656\\
100.0 & 2.298 & 0.041 & 2112.669 & 102.148 & 22.563 & 1.777 & 7270.725 & 1.649\\
110.0 & 2.309 & 0.06 & 2341.243 & 104.326 & 23.155 & 2.369 & 8220.08 & 1.641\\
120.0 & 2.323 & 0.082 & 2575.795 & 106.367 & 23.754 & 2.968 & 9190.508 & 1.632\\
130.0 & 2.34 & 0.106 & 2816.26 & 108.291 & 24.334 & 3.548 & 10180.745 & 1.623\\
140.0 & 2.361 & 0.131 & 3062.367 & 110.115 & 24.881 & 4.094 & 11189.711 & 1.613\\
150.0 & 2.384 & 0.157 & 3313.73 & 111.849 & 25.385 & 4.599 & 12216.458 & 1.603\\
200.0 & 2.54 & 0.287 & 4634.245 & 119.434 & 27.269 & 6.483 & 17589.625 & 1.56\\
150.0 & 2.384 & 0.157 & 3313.73 & 111.849 & 25.385 & 4.599 & 12216.458 & 1.603\\
298.15 & 2.964 & 0.487 & 7404.367 & 130.682 & 28.834 & 8.047 & 29079.567 & 1.503\\
300.0 & 2.973 & 0.49 & 7457.721 & 130.861 & 28.846 & 8.06 & 29306.112 & 1.503\\
350.0 & 3.224 & 0.561 & 8906.702 & 135.327 & 29.08 & 8.294 & 35547.806 & 1.485\\
400.0 & 3.488 & 0.616 & 10363.571 & 139.218 & 29.181 & 8.395 & 41997.824 & 1.473\\
450.0 & 3.761 & 0.66 & 11823.929 & 142.658 & 29.229 & 8.442 & 48630.664 & 1.463\\
500.0 & 4.039 & 0.696 & 13286.154 & 145.739 & 29.259 & 8.473 & 55426.21 & 1.455\\
600.0 & 4.609 & 0.75 & 16215.191 & 151.079 & 29.326 & 8.54 & 69443.696 & 1.444\\
700.0 & 5.191 & 0.791 & 19152.97 & 155.608 & 29.44 & 8.654 & 83952.266 & 1.437\\
800.0 & 5.782 & 0.823 & 22105.469 & 159.55 & 29.623 & 8.837 & 98882.88 & 1.43\\
900.0 & 6.381 & 0.852 & 25079.998 & 163.053 & 29.88 & 9.094 & 114184.769 & 1.425\\
1000.0 & 6.989 & 0.878 & 28083.68 & 166.217 & 30.204 & 9.418 & 129819.345 & 1.421\\
1100.0 & 7.608 & 0.903 & 31122.471 & 169.113 & 30.579 & 9.793 & 145756.449 & 1.416\\
1200.0 & 8.239 & 0.928 & 34200.737 & 171.792 & 30.99 & 10.204 & 161971.903 & 1.412\\
1300.0 & 8.883 & 0.953 & 37321.228 & 174.289 & 31.422 & 10.635 & 178445.876 & 1.408\\
1400.0 & 9.542 & 0.978 & 40485.298 & 176.634 & 31.86 & 11.074 & 195161.754 & 1.404\\
1500.0 & 10.217 & 1.003 & 43693.188 & 178.847 & 32.297 & 11.51 & 212105.359 & 1.399\\
2000.0 & 13.874 & 1.129 & 60353.995 & 188.419 & 34.278 & 13.492 & 299855.404 & 1.38\\
2500.0 & 18.083 & 1.248 & 77899.277 & 196.243 & 35.838 & 15.052 & 391921.882 & 1.364\\
3000.0 & 22.92 & 1.354 & 96138.376 & 202.89 & 37.077 & 16.291 & 487589.013 & 1.35\\
3500.0 & 28.448 & 1.45 & 114944.134 & 208.686 & 38.122 & 17.336 & 586356.362 & 1.339\\
4000.0 & 34.724 & 1.536 & 134239.943 & 213.838 & 39.043 & 18.257 & 687853.627 & 1.329\\
4500.0 & 41.805 & 1.615 & 153968.934 & 218.484 & 39.852 & 19.065 & 791795.791 & 1.321\\
5000.0 & 49.748 & 1.687 & 174065.773 & 222.719 & 40.508 & 19.722 & 897955.017 & 1.314\\
6000.0 & 68.431 & 1.809 & 214980.038 & 230.176 & 41.156 & 20.37 & 1.116191248e6 & 1.302\\
7000.0 & 91.097 & 1.899 & 256033.52 & 236.505 & 40.784 & 19.998 & 1.34130376e6 & 1.294\\
8000.0 & 117.857 & 1.954 & 296259.779 & 241.879 & 39.56 & 18.774 & 1.572254912e6 & 1.29\\
9000.0 & 148.594 & 1.977 & 334983.848 & 246.442 & 37.84 & 17.054 & 1.808162398e6 & 1.288\\
10000.0 & 182.999 & 1.973 & 371898.356 & 250.333 & 35.983 & 15.197 & 2.048285773e6 & 1.288\\
11000.0 & 220.648 & 1.95 & 406988.917 & 253.679 & 34.219 & 13.432 & 2.292017674e6 & 1.29\\
12000.0 & 261.068 & 1.914 & 440427.327 & 256.589 & 32.717 & 11.931 & 2.538869339e6 & 1.293\\
13000.0 & 303.791 & 1.871 & 472486.104 & 259.156 & 31.482 & 10.696 & 2.788452722e6 & 1.297\\
14000.0 & 348.397 & 1.825 & 503482.216 & 261.453 & 30.569 & 9.782 & 3.040462648e6 & 1.301\\
15000.0 & 394.537 & 1.78 & 533742.527 & 263.541 & 30.004 & 9.217 & 3.294660864e6 & 1.305\\
16000.0 & 441.939 & 1.736 & 563582.645 & 265.467 & 29.741 & 8.955 & 3.550862531e6 & 1.309\\
17000.0 & 490.413 & 1.697 & 593293.539 & 267.268 & 29.736 & 8.95 & 3.808925029e6 & 1.313\\
18000.0 & 539.849 & 1.664 & 623132.856 & 268.974 & 29.995 & 9.209 & 4.06873877e6 & 1.316\\
19000.0 & 590.199 & 1.636 & 653319.399 & 270.606 & 30.443 & 9.657 & 4.330219658e6 & 1.319\\
20000.0 & 641.473 & 1.613 & 684030.114 & 272.181 & 39.297 & 18.511 & 4.593302921e6 & 1.321\\
\end{tabular}
\label{tab:result-thermo-Qnorm}
\end{table*}}

\onltab{
\begin{table*}[h]
\caption{Thermodynamic properties of ortho $\rm H_2$.}
\begin{tabular}{ccccccccc} 
\hline
$T$ & $Q_{int}$ & $E_{int}/RT$ & $H-H(0)$ & $S$ & $C_p$ & $C_v$ & $-[G-H(0)]/T$ &$\gamma$\\ 

[K] & & [J] & [J mol$^{-1}$] & [J K$^{-1}$ mol$^{-1}$] & [J $K^{-1}$ mol$^{-1}$] & [J K$^{-1}$ mol$^{-1}$] & [J K$^{-1}$ mol$^{-1}$]  \\
\hline
5.0 & 3.0 & 0.0 & 103.931 & 41.756 & 20.786 & 0.0 & 63.277 & 1.667\\
10.0 & 3.0 & 0.0 & 207.861 & 56.164 & 20.786 & 0.0 & 270.633 & 1.667\\
15.0 & 3.0 & 0.0 & 311.792 & 64.592 & 20.786 & 0.0 & 532.37 & 1.667\\
20.0 & 3.0 & 0.0 & 415.723 & 70.572 & 20.786 & 0.0 & 829.423 & 1.667\\
25.0 & 3.0 & 0.0 & 519.654 & 75.21 & 20.786 & 0.0 & 1152.736 & 1.667\\
30.0 & 3.0 & 0.0 & 623.584 & 79.0 & 20.786 & 0.0 & 1496.976 & 1.667\\
35.0 & 3.0 & 0.0 & 727.515 & 82.204 & 20.786 & 0.0 & 1858.619 & 1.667\\
40.0 & 3.0 & 0.0 & 831.446 & 84.98 & 20.786 & 0.0 & 2235.16 & 1.667\\
45.0 & 3.0 & 0.0 & 935.377 & 87.428 & 20.786 & 0.0 & 2624.727 & 1.667\\
50.0 & 3.0 & 0.0 & 1039.308 & 89.618 & 20.786 & 0.0 & 3025.865 & 1.667\\
60.0 & 3.0 & 0.0 & 1247.182 & 93.408 & 20.789 & 0.003 & 3858.425 & 1.667\\
70.0 & 3.0 & 0.0 & 1455.125 & 96.613 & 20.802 & 0.016 & 4725.796 & 1.667\\
80.0 & 3.0 & 0.001 & 1663.318 & 99.393 & 20.842 & 0.056 & 5622.99 & 1.666\\
90.0 & 3.001 & 0.002 & 1872.131 & 101.853 & 20.93 & 0.144 & 6546.308 & 1.666\\
100.0 & 3.002 & 0.004 & 2082.135 & 104.065 & 21.083 & 0.297 & 7492.933 & 1.665\\
110.0 & 3.003 & 0.008 & 2294.057 & 106.085 & 21.315 & 0.529 & 8460.68 & 1.663\\
120.0 & 3.006 & 0.014 & 2508.698 & 107.952 & 21.627 & 0.84 & 9447.833 & 1.66\\
130.0 & 3.011 & 0.023 & 2726.839 & 109.698 & 22.013 & 1.227 & 10453.031 & 1.657\\
140.0 & 3.017 & 0.034 & 2949.171 & 111.346 & 22.462 & 1.676 & 11475.178 & 1.652\\
150.0 & 3.025 & 0.047 & 3176.238 & 112.912 & 22.958 & 2.171 & 12513.382 & 1.647\\
200.0 & 3.103 & 0.14 & 4390.077 & 119.878 & 25.561 & 4.775 & 17922.591 & 1.61\\
150.0 & 3.025 & 0.047 & 3176.238 & 112.912 & 22.958 & 2.171 & 12513.382 & 1.647\\
298.15 & 3.416 & 0.352 & 7069.282 & 130.738 & 28.461 & 7.675 & 29431.377 & 1.54\\
300.0 & 3.424 & 0.355 & 7121.959 & 130.914 & 28.486 & 7.7 & 29658.024 & 1.539\\
350.0 & 3.641 & 0.441 & 8559.265 & 135.345 & 28.941 & 8.155 & 35901.334 & 1.515\\
400.0 & 3.88 & 0.51 & 10011.739 & 139.223 & 29.13 & 8.344 & 42351.87 & 1.497\\
450.0 & 4.134 & 0.566 & 11470.509 & 142.66 & 29.21 & 8.424 & 48984.875 & 1.484\\
500.0 & 4.399 & 0.611 & 12932.172 & 145.74 & 29.253 & 8.467 & 55780.474 & 1.474\\
600.0 & 4.948 & 0.679 & 15860.941 & 151.079 & 29.325 & 8.539 & 69797.982 & 1.459\\
700.0 & 5.517 & 0.73 & 18798.687 & 155.608 & 29.44 & 8.653 & 84306.554 & 1.448\\
800.0 & 6.098 & 0.77 & 21751.181 & 159.55 & 29.623 & 8.837 & 99237.169 & 1.441\\
900.0 & 6.69 & 0.804 & 24725.709 & 163.053 & 29.88 & 9.094 & 114539.057 & 1.434\\
1000.0 & 7.294 & 0.835 & 27729.391 & 166.217 & 30.204 & 9.418 & 130173.634 & 1.428\\
1100.0 & 7.909 & 0.864 & 30768.183 & 169.113 & 30.579 & 9.793 & 146110.737 & 1.423\\
1200.0 & 8.537 & 0.892 & 33846.448 & 171.792 & 30.99 & 10.204 & 162326.192 & 1.418\\
1300.0 & 9.179 & 0.92 & 36966.94 & 174.289 & 31.422 & 10.635 & 178800.165 & 1.413\\
1400.0 & 9.836 & 0.948 & 40131.01 & 176.634 & 31.86 & 11.074 & 195516.042 & 1.409\\
1500.0 & 10.511 & 0.975 & 43338.899 & 178.847 & 32.297 & 11.51 & 212459.648 & 1.404\\
2000.0 & 14.173 & 1.108 & 59999.706 & 188.419 & 34.278 & 13.492 & 300209.693 & 1.383\\
2500.0 & 18.393 & 1.231 & 77544.988 & 196.243 & 35.838 & 15.052 & 392276.17 & 1.366\\
3000.0 & 23.248 & 1.34 & 95784.081 & 202.89 & 37.077 & 16.291 & 487943.302 & 1.352\\
3500.0 & 28.797 & 1.438 & 114589.782 & 208.686 & 38.122 & 17.335 & 586710.645 & 1.34\\
4000.0 & 35.096 & 1.526 & 133885.328 & 213.838 & 39.042 & 18.256 & 688207.888 & 1.331\\
4500.0 & 42.202 & 1.606 & 153613.51 & 218.484 & 39.849 & 19.063 & 792149.966 & 1.322\\
5000.0 & 50.173 & 1.678 & 173708.508 & 222.718 & 40.503 & 19.717 & 898308.966 & 1.315\\
6000.0 & 68.916 & 1.802 & 214614.043 & 230.174 & 41.144 & 20.357 & 1.116543844e6 & 1.303\\
7000.0 & 91.645 & 1.893 & 255650.45 & 236.501 & 40.76 & 19.974 & 1.34165289e6 & 1.295\\
8000.0 & 118.466 & 1.948 & 295852.244 & 241.871 & 39.533 & 18.746 & 1.572597596e6 & 1.29\\
9000.0 & 149.256 & 1.971 & 334547.337 & 246.43 & 37.81 & 17.024 & 1.808495266e6 & 1.288\\
10000.0 & 183.704 & 1.967 & 371431.338 & 250.318 & 35.952 & 15.166 & 2.048605493e6 & 1.288\\
11000.0 & 221.382 & 1.945 & 406492.024 & 253.661 & 34.191 & 13.405 & 2.292321206e6 & 1.29\\
12000.0 & 261.814 & 1.909 & 439902.437 & 256.569 & 32.688 & 11.902 & 2.53915404e6 & 1.293\\
13000.0 & 304.533 & 1.866 & 471935.664 & 259.134 & 31.454 & 10.668 & 2.788716353e6 & 1.297\\
14000.0 & 349.119 & 1.82 & 502908.834 & 261.43 & 30.545 & 9.758 & 3.040703339e6 & 1.301\\
15000.0 & 395.221 & 1.775 & 533148.787 & 263.516 & 29.974 & 9.188 & 3.294877066e6 & 1.305\\
16000.0 & 442.572 & 1.732 & 562971.041 & 265.441 & 29.735 & 8.949 & 3.551052967e6 & 1.309\\
17000.0 & 490.982 & 1.693 & 592666.488 & 267.241 & 29.721 & 8.935 & 3.809088657e6 & 1.313\\
18000.0 & 540.34 & 1.659 & 622492.742 & 268.946 & 29.99 & 9.204 & 4.068874749e6 & 1.317\\
19000.0 & 590.602 & 1.631 & 652668.607 & 270.577 & 30.434 & 9.648 & 4.330327328e6 & 1.319\\
20000.0 & 641.778 & 1.61 & 683371.061 & 272.152 & 39.29 & 18.504 & 4.59338178e6 & 1.322\\
\end{tabular}
\label{tab:result-thermo-Qortho}
\end{table*}}

\onltab{
\begin{table*}[h]
\caption{Thermodynamic properties of para $\rm H_2$.}
\begin{tabular}{ccccccccc} 
\hline
$T$ & $Q_{int}$ & $E_{int}/RT$ & $H-H(0)$ & $S$ & $C_p$ & $C_v$ & $-[G-H(0)]/T$ &$\gamma$\\ 

[K] & & [J] & [J mol$^{-1}$] & [J K$^{-1}$ mol$^{-1}$] & [J $K^{-1}$ mol$^{-1}$] & [J K$^{-1}$ mol$^{-1}$] & [J K$^{-1}$ mol$^{-1}$]  \\
\hline
5.0 & 1.0 & 0.0 & 103.931 & 32.622 & 20.786 & 0.0 & 17.605 & 1.667\\
10.0 & 1.0 & 0.0 & 207.861 & 47.03 & 20.786 & 0.0 & 179.289 & 1.667\\
15.0 & 1.0 & 0.0 & 311.792 & 55.458 & 20.786 & 0.0 & 395.355 & 1.667\\
20.0 & 1.0 & 0.0 & 415.723 & 61.437 & 20.786 & 0.0 & 646.735 & 1.667\\
25.0 & 1.0 & 0.0 & 519.654 & 66.076 & 20.786 & 0.0 & 924.377 & 1.667\\
30.0 & 1.0 & 0.0 & 623.585 & 69.865 & 20.787 & 0.001 & 1222.945 & 1.667\\
35.0 & 1.0 & 0.0 & 727.525 & 73.07 & 20.79 & 0.004 & 1538.917 & 1.667\\
40.0 & 1.0 & 0.0 & 831.508 & 75.847 & 20.806 & 0.02 & 1869.79 & 1.667\\
45.0 & 1.0 & 0.001 & 935.632 & 78.3 & 20.85 & 0.064 & 2213.703 & 1.666\\
50.0 & 1.0 & 0.002 & 1040.099 & 80.501 & 20.947 & 0.161 & 2569.224 & 1.666\\
60.0 & 1.001 & 0.009 & 1251.492 & 84.354 & 21.398 & 0.612 & 3310.871 & 1.663\\
70.0 & 1.003 & 0.025 & 1469.549 & 87.713 & 22.291 & 1.505 & 4088.379 & 1.656\\
80.0 & 1.009 & 0.054 & 1698.789 & 90.773 & 23.621 & 2.835 & 4897.857 & 1.644\\
90.0 & 1.017 & 0.097 & 1942.994 & 93.647 & 25.255 & 4.469 & 5736.931 & 1.626\\
100.0 & 1.031 & 0.151 & 2204.272 & 96.398 & 27.003 & 6.217 & 6604.1 & 1.606\\
110.0 & 1.049 & 0.215 & 2482.799 & 99.052 & 28.677 & 7.891 & 7498.282 & 1.583\\
120.0 & 1.071 & 0.283 & 2777.086 & 101.611 & 30.136 & 9.35 & 8418.531 & 1.561\\
130.0 & 1.099 & 0.354 & 3084.521 & 104.071 & 31.298 & 10.511 & 9363.887 & 1.539\\
140.0 & 1.131 & 0.423 & 3401.955 & 106.423 & 32.136 & 11.349 & 10333.31 & 1.52\\
150.0 & 1.167 & 0.488 & 3726.206 & 108.66 & 32.666 & 11.88 & 11325.683 & 1.503\\
200.0 & 1.393 & 0.727 & 5366.75 & 118.102 & 32.393 & 11.607 & 16590.726 & 1.449\\
150.0 & 1.167 & 0.488 & 3726.206 & 108.66 & 32.666 & 11.88 & 11325.683 & 1.503\\
298.15 & 1.937 & 0.892 & 8409.623 & 130.514 & 29.951 & 9.165 & 28024.137 & 1.418\\
300.0 & 1.947 & 0.894 & 8465.01 & 130.699 & 29.927 & 9.141 & 28250.377 & 1.418\\
350.0 & 2.24 & 0.919 & 9949.013 & 135.275 & 29.497 & 8.711 & 34487.222 & 1.413\\
400.0 & 2.535 & 0.933 & 11419.066 & 139.201 & 29.333 & 8.547 & 40935.684 & 1.411\\
450.0 & 2.831 & 0.944 & 12884.19 & 142.653 & 29.283 & 8.497 & 47568.031 & 1.409\\
500.0 & 3.128 & 0.951 & 14348.103 & 145.738 & 29.278 & 8.492 & 54363.419 & 1.408\\
600.0 & 3.725 & 0.963 & 17277.941 & 151.079 & 29.329 & 8.542 & 68380.839 & 1.406\\
700.0 & 4.324 & 0.973 & 20215.82 & 155.608 & 29.44 & 8.654 & 82889.401 & 1.404\\
800.0 & 4.928 & 0.983 & 23168.333 & 159.55 & 29.623 & 8.837 & 97820.014 & 1.403\\
900.0 & 5.536 & 0.994 & 26142.863 & 163.053 & 29.88 & 9.094 & 113121.903 & 1.401\\
1000.0 & 6.151 & 1.006 & 29146.545 & 166.217 & 30.204 & 9.418 & 128756.479 & 1.399\\
1100.0 & 6.774 & 1.019 & 32185.337 & 169.113 & 30.579 & 9.793 & 144693.583 & 1.397\\
1200.0 & 7.406 & 1.034 & 35263.602 & 171.792 & 30.99 & 10.204 & 160909.037 & 1.395\\
1300.0 & 8.051 & 1.051 & 38384.094 & 174.289 & 31.422 & 10.635 & 177383.01 & 1.392\\
1400.0 & 8.709 & 1.069 & 41548.164 & 176.634 & 31.86 & 11.074 & 194098.888 & 1.389\\
1500.0 & 9.382 & 1.089 & 44756.054 & 178.847 & 32.297 & 11.511 & 211042.494 & 1.386\\
2000.0 & 13.015 & 1.193 & 61416.861 & 188.419 & 34.278 & 13.492 & 298792.538 & 1.371\\
2500.0 & 17.181 & 1.299 & 78962.144 & 196.243 & 35.838 & 15.051 & 390859.016 & 1.357\\
3000.0 & 21.964 & 1.397 & 97201.262 & 202.89 & 37.077 & 16.291 & 486526.149 & 1.345\\
3500.0 & 27.428 & 1.486 & 116007.188 & 208.686 & 38.122 & 17.336 & 585293.51 & 1.335\\
4000.0 & 33.632 & 1.568 & 135303.791 & 213.838 & 39.047 & 18.26 & 686790.847 & 1.326\\
4500.0 & 40.634 & 1.644 & 155035.207 & 218.485 & 39.859 & 19.073 & 790733.265 & 1.318\\
5000.0 & 48.493 & 1.713 & 175137.568 & 222.721 & 40.522 & 19.736 & 896893.167 & 1.311\\
6000.0 & 66.995 & 1.831 & 216078.024 & 230.183 & 41.192 & 20.405 & 1.11513346e6 & 1.3\\
7000.0 & 89.472 & 1.919 & 257182.729 & 236.52 & 40.847 & 20.061 & 1.340256372e6 & 1.292\\
8000.0 & 116.05 & 1.972 & 297482.384 & 241.903 & 39.641 & 18.855 & 1.57122686e6 & 1.288\\
9000.0 & 146.624 & 1.994 & 336293.379 & 246.476 & 37.931 & 17.145 & 1.807163794e6 & 1.286\\
10000.0 & 180.9 & 1.99 & 373299.41 & 250.377 & 36.071 & 15.285 & 2.047326615e6 & 1.287\\
11000.0 & 218.462 & 1.966 & 408479.598 & 253.731 & 34.302 & 13.516 & 2.29110708e6 & 1.288\\
12000.0 & 258.842 & 1.93 & 442001.997 & 256.649 & 32.799 & 12.012 & 2.538015238e6 & 1.292\\
13000.0 & 301.576 & 1.887 & 474137.426 & 259.222 & 31.549 & 10.763 & 2.787661831e6 & 1.295\\
14000.0 & 346.243 & 1.84 & 505202.36 & 261.525 & 30.652 & 9.865 & 3.039740575e6 & 1.299\\
15000.0 & 392.49 & 1.794 & 535523.749 & 263.617 & 30.079 & 9.293 & 3.294012259e6 & 1.304\\
16000.0 & 440.045 & 1.75 & 565417.46 & 265.546 & 29.776 & 8.99 & 3.550291223e6 & 1.308\\
17000.0 & 488.713 & 1.711 & 595174.689 & 267.35 & 29.787 & 9.0 & 3.808434146e6 & 1.311\\
18000.0 & 538.38 & 1.676 & 625053.198 & 269.058 & 30.023 & 9.237 & 4.068330833e6 & 1.315\\
19000.0 & 588.994 & 1.648 & 655271.777 & 270.692 & 30.442 & 9.656 & 4.329896651e6 & 1.318\\
20000.0 & 640.561 & 1.625 & 686007.275 & 272.268 & 39.406 & 18.62 & 4.593066343e6 & 1.32\\
\end{tabular}
\label{tab:result-thermo-Qpara}
\end{table*}}


\begin{acknowledgements}
We thank \AA ke Nordlund and Tommaso Grassi for a critical reading of this manuscript and all the very valuable discussions. We are also grateful to the referee, Robert Kurucz, for valuable comments and suggestions. AP work is supported by grant number 1323-00199A from the Danish Council for Independent Research (FNU).
\end{acknowledgements}

\end{document}